\documentclass[%
 aip,
 amsmath,amssymb,
 reprint,%
]{revtex4-1}

\usepackage{graphicx}
\usepackage{dcolumn}
\usepackage{bm}

\usepackage[utf8]{inputenc}
\usepackage[T1]{fontenc}
\usepackage{mathptmx}

\begin{document}

\preprint{AIP/123-QED}

\title[Nonlinear flexural-gravity waves]{Nonlinear flexural-gravity waves due to a body submerged in the uniform stream.} 

\author{Y.A. Semenov}
\altaffiliation[]{}

\affiliation{Institute of Hydromechanics of the National Academy of Sciences of Ukraine, 8/4 Maria Kapnist Street, 03680 Kiev, Ukraine} 

\date{\today}

\begin{abstract}
The two-dimensional nonlinear problem of steady flow past a body submerged beneath an elastic sheet is considered. The mathematical model is based on the velocity potential theory with fully nonlinear boundary conditions on the fluid boundary and on the elastic sheet, which are coupled throughout the numerical procedure. The integral hodograph method is employed to derive the complex velocity potential of the flow which contains the velocity magnitude on the interface in explicit form. The coupled problem has been reduced to a system of nonlinear equations with respect to the unknown magnitude of the velocity on the interface, which is solved using a collocation method. Case studies are undertaken for both subcritical and supercritical flow regimes. Results for interface shape, bending moment and pressure distribution are presented for the wide ranges of Froude numbers and depths of submergence. According to the dispersion equation, two waves on the interface may exist. The first, longest wave is that caused by gravity, and the second, shorter wave is that caused by the elastic sheet. The obtained solution exhibits strongly nonlinear interaction of these waves above the submerged body. It is found that near the critical Froude number, there is a range of submergences in which the solution does not converge.
\end{abstract}

\maketitle

\section{\label{sec:level1}INTRODUCTION }
The problem of interaction between a fluid and an elastic boundary is a classical problem of fluid mechanics which is of interest in offshore and polar engineering, medicine and many industrial applications. In the last two decades, this topic has received much attention due to the melting of ice in the Arctic regions, opening new routes for ships and new regions for resource exploration \cite{Squire_1988,Squire_1996,Korobkin_2011}. Most of the studies are devoted to wave propagation along the ice sheet, its response on a moving load, and effects of heterogeneous properties of the ice sheet such as a floe, polynya, cracks, etc. \cite{Guyenne_2012,Guyenne_2012,Guyenne_2017,Li_2018}.

	The works studying the interaction between the flow bounded by the ice sheet and the body submerged in the fluid started relatively recently. Das and Mandal \cite{Das_2006} considered oblique wave scattering by a circular cylinder submerged beneath a uniform ice sheet of infinite extent and determined the transition and reflection coefficients. To solve the problem, they employed the multipole expansion method. Sturova \cite{Sturova_2015} applied the method of matched eigenfunction expansions and studied the interaction of a submerged cylinder and an inhomogeneous ice sheet including a floe or polynya. Tkacheva \cite{Tkacheva_2015} considered oscillations of a cylindrical body submerged in a fluid beneath an ice sheet and solved the problem through the Wiener-Hopf technique. Savin and Savin \cite{Savin_2012} considered the ice perturbation by a dipole submerged in water of infinite depth. They applied the complex variable technique and the Fourier transform to solve the Laplace equation. Shishmarev at al. \cite{Shishmarev_2019}  studied the strains in an ice cover of a frozen channel which are caused by a three-dimensional dipole moving under the ice at a constant speed. Li et al. \cite{Li_2019} considered a circular cylinder submerged below an ice sheet in water of finite depth. The solution method is based on the derived Green function which satisfies the boundary conditions on the ice/water interface. All the works mentioned above regarding submerged bodies are based on linear potential flow theory, and the boundary value problem is usually formulated in the frequency domain.

	The nonlinear theory of hydroelasticity is currently under development. The unknown shape of the ice/fluid interface and its higher-order derivatives, which have to satisfy the dynamic boundary condition, are the main challenge to derive analytical solutions or develop computational approaches. As the dynamic boundary condition gets more complicates, e.g. include gravity, surface tension and/or elasticity of the sheet covering the fluid, it increases the level of the mathematical challenge which has to be addressed. The simplest form of the dynamic boundary condition corresponds to free streamline flows for which the velocity magnitude on the free streamline is assumed to be constant. This class of flows is well developed and presented in classical books by Milne-Thomson
\cite{Milne-Thomson}, Birkhoff and Zarantonello \cite{Birkhoff}, and Gurevich \cite{Gurevich}.

For free-surface flows, gravity leads to an additional term in the dynamic boundary condition which relates the velocity magnitude and the vertical coordinate of the free surface. This kind of problem can be reduced to a singular integrodifferential equation whose form depends on the solution method and the choice of the governing functions. Various forms of the integro-differential equation were derived by Forbes and Schwartz \cite{Forbes_1982}, King and Bloor \cite{King_Bloor}, and Faltinsen and Semenov \cite{Faltinsen_Semenov}.

For capillary free-surface flows, the dynamic boundary condition comprises the curvature of the free surface which involves the first and second derivatives of the free surface. Fewer analytical solutions for purely capillary waves are presented in literature. Crapper \cite{Crapper} developed a closed-form solution for a fluid of infinite depth, and Kinnersley \cite{Kinnersley} extended his method to a fluid sheet. Crowdy \cite{Crowdy} developed a method based on complex function theory and retrieved Kinnersley's solution in much simpler form and obtained new solutions for steady capillary waves on a fluid annulus. However, the extension of the method to fluid/structure interaction problems with surface tension seems to be nontrivial. Alternatively, several numerical methods have been developed to solve the capillary and capillary-gravity flows. Schwartz and Vanden-Broeck \cite{Schwartz_Vanden-Broeck} proposed a method based on a boundary-integral formulation and finite difference approximation of the derivatives and applied it to the purely capillary and capillary-gravity waves. Vanden-Broeck and Miloh \cite{Vanden-Broeck_Miloh} proposed numerical methods based on the Fourier-series expansion and studied steep gravity waves. Later, the method was adopted by Blyth and Vanden-Broeck \cite{Blyth_Vanden-Broeck} and Blyth and P\u{a}r\u{a}u \cite{Blyth_Parau} to compute nonlinear capillary waves on fluid sheets of finite thickness. Yoon and Semenov \cite{Yoon_Semenov} considered a cavity flow past a circular cylinder in the presence of surface tension and derived the solution using the integral hodograph method. They derived a singular integral equation with respect to the velocity magnitude, which is solved by the method of successive approximations. The method can be applied to solve problems with a more complicated form of the dynamic boundary condition which comprises higher-order derivatives of the free surface. However, the higher-order derivatives of the interface which appear in the dynamic boundary condition results to a higher-order hypersingular integral equation. A special numerical treatment is required to solve this type of integral equation.

The nonlinear theory of hydroelastic waves, for which the dynamic boundary condition gets more complicated, has been studied intensively in recent decades with emphasis on waves generated by a moving load. Most of the studies are focused on the analysis and simulation of hydroelastic waves, which account for the nonlinearity of the potential flow and elastic sheet deformations. ParauP\u{a}r\u{a}u and Dias \cite{Parau_Dias} derived a forced nonlinear Schrodinger equation for ice sheet deflection and studied the weakly nonlinear effects. Fully nonlinear computations based on the boundary integral method were presented by Guyenne and P\u{a}r\u{a}u \cite{Guyenne_2012}. The nonlinear response of an infinite ice sheet in the time domain has been studied by Bonnefoy et al. \cite{Bonnefoy} using a higher-order spectral method. They found that at a critical speed at which the linear response is infinite, the nonlinear solution remains bounded. Despite the progress in the development of numerical methods to solve nonlinear problems of hydroelastic waves, their extension to a body submerged in fluid beneath an elastic sheet seems to be not easy, since the flow potential has to satisfy an additional boundary condition on the body surface.

	In the present paper, we study a fully nonlinear problem of the hydroelastic waves generated by a body submerged in the fluid beneath an elastic sheet. We shall use the model of potential flow with the fully nonlinear kinematic and dynamic boundary conditions on the submerged rigid body and the elastic sheet which is modelled using the Cosserat theory of hyperelastic shells suggested by Plotnikov and Toland \cite{Plotnikov}. To solve the nonlinear problem, we adopt the solution for a cylindrical body moving beneath a free surface \cite{Semenov_Wu}, which is obtained using the integral hodograph method. An expression for the flow potential which includes the velocity magnitude on the free surface in an explicit form has been derived. This gives the possibility to adopt the solution to in  the present problem, because the velocity magnitude on the interface between the fluid and elastic sheet appears in the dynamic boundary condition explicitly. The coupling of the elastic sheet and fluid problems is based on the condition of the same pressure on the interface, one from flow dynamics and the second from elastic sheet equilibrium.

	The derivation of the flow potential which contains in explicit form the velocity magnitude on the interface and the numerical method to solve the coupled fluid/elastic sheet interaction problem are presented in Section 2. The extended numerical results are discussed in Section 3.

\section{\label{sec:leve21}THEORETICAL ANALYSIS}
We consider a two-dimensional steady flow past a cylindrical body submerged beneath an elastic sheet which is modelling the ice cover. The characteristic length of the body is $L$, and the thickness of the sheet is $B_i$. A Cartesian coordinate system $XY$ is defined with the origin at a point inside the body and the $X$-axis along the velocity direction of the incoming flow with a constant speed $U$. The $Y-$axis points vertically upwards. The fluid is assumed to be inviscid and incompressible, and the flow is irrotational. The elastic sheet is modelled by the Cosserat theory of hyperelastic shells \cite{Plotnikov}. The submerged rigid body is assumed to have an arbitrary shape which can be defined by the slope of the body as a function of the arc length coordinate $S$, or $\beta_b=\beta_b(S)$. The interface between the elastic sheet and the liquid is defined by function $Y(X)$. The interactions of the submerged body, flow and the elastic sheet may generate waves extending to infinity in both upstream and downstream directions. On the other hand, the flow is uniform at infinity, $Y\rightarrow \infty$ and $-\infty<X<\infty$, and the velocity is constant there. In order to provide the same value of the flow velocity at infinity in all directions, we introduce damping regions $P_1P_2$ and $T_1T_2$  upstream and downstream, respectively, where a term providing the wave damping is added in the dynamic boundary condition to provide the same velocity magnitude at points $P_2$ and $T_2$, or $V_{P2}=V_{T2}=U$. Outside the interval $P_2T_2$, the flow velocity on the interface $V(X)\equiv U$  including infinity. Thus, the fluid surface has a limit $Y(x)_{|x|\rightarrow \infty}=H$ which is defined as the submergence of the cylinder measured from the origin of the coordinate system.
\begin{figure*}
\includegraphics{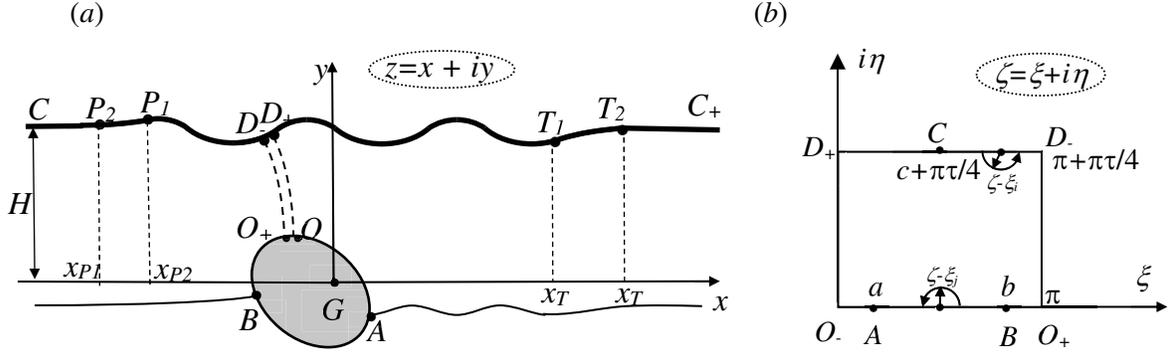}
\caption{\label{fig:wide} ($a$) Physical plane and ($b$) the parameter $\zeta-$plane. }
\end{figure*}

We will derive the complex potential of the flow, $W=W(Z)$, with $Z=X+iY$. For the steady flow, the kinematic conditions on the body surface and the interface mean that the stream function is constant, or $\Im\{W(Z)\}=const.$, as they both are streamlines. The dynamic boundary condition on the interface is obtained from the Bernoulli equation assuming that the hydrodynamic pressure on the interface is the same as the pressure conditioning the bending of the elastic sheet.
\begin{equation}
\label{eqv1} \rho \frac{V^2}{2} + \rho gY + P_{ice}=\rho \frac{U^2}{2} + \rho gH + P_a,
\end{equation}
where $U$ is the speed of the incoming flow, $\rho$ is the liquid density, $V=|dW/dZ|$  is the magnitude of the complex velocity, $g$ is the acceleration due to gravity, $H$ is the depth of submergence, and $P_a$ is the atmospheric pressure. The pressure due to the bending of the elastic sheet is \cite{Guyenne_2012,Plotnikov}
\begin{equation}
\label{eqv2} P_{ice}=D^{\prime}B^3_i\left( \frac{d^2\kappa}{dS^2}+\frac{1}{2}\kappa^3\right) + P_a,
\end{equation}
where $D^{\prime}=E/(12(1-\nu))$ is the flexural rigidity of the elastic sheet, $\kappa$ is the curvature of the interface, $B_i$ is the thickness of the elastic sheet. Equation (\ref{eqv2}) corresponds to the assumptions: the elastic sheet is inextensible and not prestressed \cite{Blyth_Parau}.

Two different Froude numbers can be defined based on the characteristic length $L$ or the depth of submergence $H$, respectively:
\begin{equation}
\label{eqv3} F=\frac{U}{\sqrt{gL}}, \qquad   F_h=\frac{U}{\sqrt{gH}}.
\end{equation}
Using nondimensionalisation based on $U$, $L$ and $\rho$, we have $v=V/U$, $x=X/L$, $y=Y/L$, $h=H/L$, $s=S/L$, $b_i=B_i/L$, and $W(Z)=ULw(z)$. Replacing   in equation (\ref{eqv1}) by (\ref{eqv2}), the dynamic boundary condition (\ref{eqv1}) takes the form
\begin{equation}
\label{eqv4} v^2=1-\frac{2(y-h)}{F^2}-2D\left( \frac{d^2\kappa}{dS^2}+\frac{1}{2}\kappa^3\right),
\end{equation}
where \[D=\frac{D^\prime}{\rho gLF^2},  \qquad  \kappa=\frac{d\delta}{ds}. \] The angle $\delta=\arcsin(dy/ds)=\pi+\beta$
is the angle between the $X-$axis and the unit tangential vector $\bold{\tau}$  which is opposite to the velocity direction $\beta$.
The normal vector $\bold{n}$ is directed from the liquid region outwards, while along the interface the spatial coordinate   increases in the direction of the vector  $\bold{\tau}$ such that the liquid region is on the left (see Fig. \ref{fig:wide}a).

Equation (\ref{eqv4}) contains the velocity magnitude along the interface, the wave elevation $y$ and its derivatives which will be related in the following throughout the derived expression for the flow potential.

\subsection{\label{sec21} Hodograph method.}
Finding the function $w=w(z)$  directly is a complicated problem since the boundary of the flow region is unknown. Instead, Joukovskii \cite{Joukovskii_1890} and Michell \cite{Michell} proposed to introduce an auxiliary parameter plane, or $\zeta-$plane, which was typically chosen as the upper half-plane. Then, they considered two functions, which were the complex potential $w$ and the function $\omega=-\ln(dw/dz)$, both as functions of the parameter variable $\zeta$. When $w(\zeta)$  and $\omega(\zeta)$  are derived, the velocity and the flow region can be obtained in the parameter form as follows:
\begin{equation}
\label{eqv5} \frac{dw}{dz}=\exp[-\omega(\zeta)],  \qquad  z(\zeta)=z_0+\int_0^\zeta\frac{dw}{d\zeta^\prime}/\frac{dw}{dz}d\zeta^\prime,
\end{equation}
where the function $z=z(zeta)$ is called as the mapping function.

The flow region beneath the interface and outside the body is a doubly connected domain. A canonical region of a doubly connected domain is an annulus. By making a cut connecting the external and the internal circles of the annulus, the doubly connected region becomes simply connected. As shown in Fig. \ref{fig:wide}a, $O_-D_+$  and $O_+D_-$  are the two sides of the cut which could have an arbitrary shape but form a right angle with the flow boundary at both the body surface (points $O_-$ and $O_+$) and the liquid/ice interface (points $D_-$ and $D_+$).

The simply connected flow region $C_-D_-O_+BAO_-D_+C_+$ is then transformed into the rectangular domain $O_-D_+CD_-BAO_+$ in the parameter plane. An upper half-plane or unit circle is usually chosen as the parameter plane. However, the flow region may have corner points at which the mapping $z(\zeta)$ will be not conformal. Chaplygin (see chapter 1(5) in the book \cite{Gurevich}) pointed out that there is flexibility in the choice of the region of the parameter variable. It should be composed of straight lines and arcs of circles in such a way that by means of image transformations of these regions it is possible to cover the whole complex plane in a simple manner.

When solving boundary problems, the shape of the auxiliary parameter region is usually chosen with the aim of obtaining the solution of a problem in the simplest form with a minimal number of singular points at which the transformation of the parameter region onto the complex potential region, $w$, and the region of the  function $dw/dz$ is not conformal. In the present case of the doubly connected flow region, the additional corner points appear at the intersections of two sides of the cut and the flow boundary. In order to provide a conformal mapping at these corner points, we have chosen the rectangle as the parameter domain, which also has right angles at points $O_-, O_+, D_-, D_+$, e.g. the same angles as in the physical plane.

When the parameter region is chosen as a half-plane or the first quadrant, the polynomial functions are usually used to construct the mapping function  $z(\zeta)$. Here, for the rectangular domain, the polynomial functions will be replaced by Jacobi's theta functions \cite{Birkhoff}, which are quasi-doubly-periodic functions. Due the periodicity, they naturally satisfy the same conditions on both sides of the cut. Jacobi's functions have been used to solve free surface problems involving doubly connected flow regions in the books \cite{Milne-Thomson, Birkhoff, Gurevich, Terentiev}.

We may choose the coordinates of the rectangle vertices $O_-O_+D_-D_+$ as $(0,0)$, $(\pi,0)$, $(\pi,\pi\tau/4)$ and $(0,\pi\tau/4)$, respectively, as shown in Fig. \ref{fig:wide}a. Here, $\tau$ is an imaginary number. The horizontal length of the rectangle is then equal to $\pi$, and its vertical length is equal to $\pi|\tau|/4$.

In the flow region, there are two stagnation points marked as $A$, where two streamlines merge into one, and $B$, where a streamline splits into two branches. Positions of these points in parameter plane $\zeta=a$, $\zeta=b$  as well as the position of point $C$, $\zeta=c+\pi\tau/4$ which corresponds to infinity in the physical plane.  The parameters $a$, $b$ and $c$ should be determined from additional conditions at the solution of the problem.

The interval $0\leq \xi \leq \pi$ on the real axis corresponds to the body boundary. The interval $c< \xi \leq\pi$, $i\eta=\pi\tau/4$ corresponds to part of the interface $D_-C_-$, and the interval  $0\leq\xi< c$, $i\eta=\pi\tau/4$ corresponds to the other part of the interface $D_+C_+$. It should be noticed that points $C_-$ at $x\rightarrow -\infty$ and $C_+$ for $x\rightarrow +\infty$ in the physical plane have been transformed to the same point $C$ in the parameter region $\zeta$.

\subsection{\label{sec22} Integral hodograph method:derivation of the governing functions $dw/dz$  and $dw/d\zeta$.}

At this stage we denote the angle of the velocity direction along the body as $\beta_b(\xi)$ and the velocity magnitude on the free surface as $v(\xi)$. With these notations, we have the following boundary-value problem for the function of complex velocity, $dw/dz$:
\begin{equation}
\label{eqv6}
\left|\frac{dw}{dz} \right|=v(\xi), \qquad 0\leq \xi {\leq}\pi, \quad \eta=\pi\tau/4.
\end{equation}

\begin{equation}
\label{eqv7}
\chi(\xi ) = \arg \left(\frac{dw}{dz} \right) = \left\{{\begin{array}{l}
- \beta_b(\xi),\qquad \quad 0 \leq \xi < a, \quad \eta=0, \\
- \beta_b(\xi) - \pi,\quad \,\, a < \xi < b, \quad \eta=0, \\
- \beta_b(\xi) - 2\pi,\quad    b < \xi \leq \pi, \quad \eta=0.
\end{array}} \right.
\end{equation}
\begin{equation}
\label{eqv8}
\frac{dw}{dz}(\xi=0, i\eta)  = \frac{dw}{dz}(\xi=\pi, i\eta), \qquad 0\leq i\eta \leq \pi\tau/4.
\end{equation}
In (\ref{eqv7}) the argument of complex velocity has the jumps equal to $-\pi$ at stagnation points $A$ ($\zeta=a$) and $B$ ($\zeta=b$) due to the jump of the velocity direction when passing through the stagnation point.  The two vertical sides of the rectangle in the parameter plane correspond to the two sides of the cut in the physical plane. The velocities on both sides of the cut are the same and therefore the condition of periodicity can be applied on the vertical sides the rectangle. The solution of the boundary value problem (\ref{eqv6})-(\ref{eqv8}) can be obtained by applying the integral formulae derived in \cite{Sem_Wu2020},
\begin{eqnarray}
\label{eqv9}
\frac{dw}{dz}&=&v(\pi)\exp\left[-\frac{1}{\pi}\int_0^\pi\frac{d\chi}{d\xi}\ln\left(\frac{\vartheta_1(\zeta-\xi)}{\vartheta_1(\zeta-\xi-\pi\tau/2)} \right)d\xi\right. \nonumber \\
&+&\frac{i}{\pi}\left.\int_\pi^0\frac{d\ln v}{d\xi}\ln\left(\frac{\vartheta_1(\zeta-\xi-\pi\tau/4)}{\vartheta_1(\zeta-\xi+\pi\tau/4)} \right)d\xi +i\chi(\pi)\right].
\end{eqnarray}
It can be easily verified that for $0<\xi<\pi$, $\eta=0$ the argument, $\arg[(dw/dz)_{\zeta=\xi,\eta=0}]=\chi(\xi)$, while for $0<\xi<\pi$, $i\eta=\pi\tau/4$, the modulus $|dw/dz|_{\zeta=\xi,i\eta=\pi\tau/4}=v(\xi)$, i.e. the boundary conditions (\ref{eqv6}) and (\ref{eqv7}) are satisfied. The boundary condition (\ref{eqv8}) is satisfied due to periodicity of the function $\vartheta_1(\zeta)$. By substituting the boundary conditions (\ref{eqv6}) and (\ref{eqv7}) into (\ref{eqv9}) and evaluating the first integral over the step change in the function $\chi(\xi)$ at points $\zeta=a$  and $\zeta=b$, we obtain the expression for the complex velocity in the rectangle $O_-, O_+, D_=, D_+$ \cite{Sem_Wu2020},
\begin{eqnarray}
\label{eqv10}
\frac{dw}{dz}&=&v_D\frac{\vartheta_1(\zeta-a)\vartheta_1(\zeta-b)}{\vartheta_4(\zeta-a)\vartheta_4(\zeta-b)}\exp\left[\frac{1}{\pi}\int_0^\pi
\frac{d\beta_b}{d\xi}\ln\frac{\vartheta_1(\zeta-\xi)}{\vartheta_4(\zeta-\xi)}d\xi\right. \nonumber \\
&+&\frac{i}{\pi}\left.\int_\pi^0\frac{d\ln v}{d\xi}\ln\frac{\vartheta_1(\zeta-\xi-\pi\tau/4)}{\vartheta_4(\zeta-\xi-\pi\tau/4)}d\xi-i\beta_O\right].
\end{eqnarray}
where $\beta_{O}$  is the angle at point $O_-$  which is zero if point $O_-$ is  the highest point of the body. The constant $v_D$ or the velocity magnitude at point $D_+$, is determined by satisfying the velocity at infinity, $\zeta=c+\pi\tau/4$, which is $1$, as it has been chosen as the reference velocity, or
\begin{equation}
\label{eqv11}
\left|\frac{dw}{dz}\right|_{\zeta=c+\pi\tau/4}=1
\end{equation}

For steady flows, the stream function $\psi=\Im(w)$  takes constant values along the body and the interface. According to Chaplygin's special point method \cite{Gurevich}, to determine the function $w=w(\zeta)$, it is sufficient to analyse all special points where the mapping is not conformal. These are the stagnation points $A (\zeta=a)$ and $B (\zeta=b)$ and point $C (\zeta=c+\pi\tau/4)$  corresponding to infinity in the $w-$plane. The order of the function $w-w(\zeta)$ at these points can be determined by analysing the behaviour of the argument of $w(\zeta)$ in the vicinity of these points. Then, the derivation of the complex potential can be obtained in the form \cite{Sem_Wu2020}.
\begin{eqnarray}
\label{eqv12}
\frac{dw}{d\zeta}=K\frac{\vartheta_1(\zeta-a)\vartheta_4(\zeta-a)\vartheta_1(\zeta-b)\vartheta_4(\zeta-b)} {\vartheta_1^2(\zeta-c-\pi\tau/4)\vartheta_1^2(\zeta-c+\pi\tau/4)}.
\end{eqnarray}

Dividing (\ref{eqv12}) by (\ref{eqv10}), we obtain the derivative of the mapping function as
\begin{eqnarray}
\label{eqv13}
\frac{dz}{d\zeta}&=&\frac{K}{v_D}\frac{\theta_4^2(\zeta-a)\theta_4^2(\zeta-b)}{\theta_1^2(\zeta-c-\pi\tau/4)\theta_1^2(\zeta-c+\pi\tau/4)}\\ \nonumber
&&
\times\exp\left[-\frac{1}{\pi}\int_0^\pi \frac{d\beta_b}{d\xi}\ln\frac{\theta_1(\zeta-\xi)}{\theta_4(\zeta-\xi)}d\xi \right. \\ \nonumber
&-&\left. \frac{i}{\pi} \int_\pi^0\frac{d\ln v}{d\xi}\ln\frac{\theta_1(\zeta-\xi-\pi\tau/4)}{\theta_4(\zeta-\xi-\pi\tau/4)}d\xi+i\beta_O\right].
\end{eqnarray}
whose integration along the intervals $0\leq\xi<c$ and  $c<\xi\leq\pi$ at $\eta=\pi\tau/4$ in the $\zeta-$plane provides the parts $D_+C_+$  and $D_-C_-$ of the free surface $C_-C_+$ in $\zeta-$plane, respectively. The parameters $a, b, c, \tau$ and $K$, and the functions $\beta_b(\xi)$ and $d(\ln v)/d\xi$ have to be determined from physical considerations and the kinematic boundary condition on the body surface and the dynamic boundary conditions on the free surface.

\subsection{\label{sec23} System of equations for parameters $a, b, c, \tau$  and $K$.}

At infinity, point $C_-C_+ (\zeta=c+\pi\tau/4)$, the velocity approaches unit (since this velocity is chosen as the reference velocity), and its direction is along the X-axis. Therefore, the argument of the complex velocity (10) at point $\zeta_C=c+\pi\tau/4$ should be equal to zero
\begin{equation}
\label{eqv14}
\arg\left(\frac{dw}{dz}\right)_{\zeta=\zeta_C}=0.
\end{equation}
The scale factor $K$ is determined by the length $S_b$ which is the perimeter of the body cross-section
\begin{equation}
\label{eqv15}
\int_0^\pi\frac{ds_b}{d\xi}d\xi=S_b.
\end{equation}
where
\[\frac{ds_b}{d\xi}=\left|\frac{dz}{d\zeta}\right|_{\zeta=\xi}.\]
The free surface on the left and right hand sides at infinity has the same value of $y-$coordinate. This is also equivalent to that the stream function $\psi=\Im(w)$ is continuous across the cut, or $\Im(w_{D_-})-\Im(w_{D_+})=0$. By integrating $\Im(dw/d\zeta)$ along $D_-D_+$ passing the point $\zeta_C$ along a semi-circle $C^\prime$ of an infinitesimal radius $\varepsilon$, at which $dw/d\zeta$ in Eq.(\ref{eqv12}) has the second order singularity, we have
\begin{eqnarray}
\label{neq1}
&&\Im\left(\int_\pi^{c+\varepsilon}\frac{dw}{d\zeta}d\zeta + \oint_{C^\prime}\frac{dw}{d\zeta}d\zeta + \int_{c-\varepsilon}^0\frac{dw}{d\zeta}d\zeta\right)  \nonumber \\
&=& \Im\left(\oint_{C^\prime}\frac{dw}{d\zeta}d\zeta\right) = \Im\left(i\pi
\begin{array}{c}
~~~  \\
\mbox{Res}\\
~^{\zeta=\zeta_C}
\end{array}
\frac{dw}{d\zeta}\right) \\
&=&\Im\left\{i\pi \frac{d}{d\zeta}\left[\frac{dw}{d\zeta}(\zeta-\zeta_C)^2\right]_{\zeta=\zeta_C}\right\}.
\nonumber
\end{eqnarray}
Here the first and third terms on the left hand are zero because $\Im(w)=const.$ on the free surface. From this equation it follows
\begin{equation}
\label{eqv16}
a+b=2c.
\end{equation}

The depth of submergence, $h$, and the flowrate, $Q$, between the body and the free surface are related. Therefore, instead of a condition for the depth $h$, we can use the following condition for the given flowrate $Q$, which is the integral of the derivative of the complex potential along the side $O_-D_+$
\begin{equation}
\label{eqv17}
\Im\left(\int_0^{\pi\tau/4}\frac{dw}{d\zeta}d\zeta\right)=Q.
\end{equation}

We may place a vortex with circulation $\Gamma$ at the centre of the cylinder, which can be nondimensionalized as $\gamma=\Gamma/(2\pi UL)$. For a circular cylinder, this does not affect the impermeable body surface boundary condition, but does change the positions of the stagnation points and also affects the free surface boundary. For a hydrofoil, $\gamma$ should be determined through the Kutta condition at the trailing edge.

Integrating $dw/d\zeta$ along the body surface in the parameter plane, we have
\begin{equation}
\label{eqv18}
\Re\left(\int_0^{\pi}\frac{dw}{d\zeta}d\zeta\right)=2\pi\gamma.
\end{equation}
In the case $\gamma\neq0$, the real part of the potential, $\phi=\Re(w)$, have a jump on the sides $O_-D_-$ and $O_+D_+$ of the cut, while the complex velocity,  $dw/dz$ and the stream function $\psi=\Im(w)$  are still continuous across the cut.

Equations (\ref{eqv14}) - (\ref{eqv18}) allow us to determine the unknown parameters $a, b, c, \tau$ and $K$, which appear in the governing equations (\ref{eqv10}), (\ref{eqv12}) and (\ref{eqv13}), once the functions $v(\xi)$ and $\beta_b(\xi)$ are specified.

\subsection{\label{sec24} Kinematic boundary conditions on the body surface.}
By integrating the modulus of the derivative of the mapping function (\ref{eqv13}) along the side $O_-O_+$ in the parameter plane, we can obtain the spatial coordinate along the body as a function of the parameter variable
\begin{equation}
\label{eqv19}
s_b(\xi)=\int_0^\xi\frac{ds_b}{d\xi^\prime}d\xi^\prime,
\end{equation}
where $ds_b/d\xi=|dz/d\zeta|_{\zeta=\xi, \eta=0}$. Since the function $\beta_b(s_b)$ is known, the function $\beta_b(\xi)$ can be  determined from the following integro-differential equation:
\begin{equation}
\label{eqv20}
\frac{d\beta_b}{d\xi}=\frac{d\beta_b}{ds_b}\frac{ds_b}{d\xi}.
\end{equation}
By substituting $dz/d\zeta$ from (\ref{eqv13}), this equation takes the form
\begin{eqnarray}
\label{eqv21}
\frac{d\beta_b}{d\xi}&=&\kappa[s_b(\xi)]\frac{K}{v_D}\left|\frac{\theta_4^2(\xi-a)\theta_4^2(\xi-b)}{\theta_1^2(\xi-c-\pi\tau/4)\theta_1^2(\xi-c+\pi\tau/4)} \right| \nonumber \\
&\times& \exp\left[-\frac{1}{\pi}\int_0^\pi \frac{d\beta_b}{d\xi^\prime}\ln\frac{\theta_1(\xi-\xi^\prime)}{\theta_4(\xi-\xi^\prime)}d\xi^\prime \right. \nonumber \\
&-& \left. \frac{i}{\pi} \int_\pi^0\frac{d\ln v}{d\xi^\prime}\ln\frac{\theta_1(\xi-\xi^\prime-\pi\tau/4)}{\theta_4(\xi-\xi^\prime-\pi\tau/4)}d\xi^\prime\right],
\end{eqnarray}
where $\kappa(s_b)=d\beta_b/ds_b$  is the curvature of the body.

\subsection{\label{sec25} Nonlinear dynamic boundary condition.}
The dynamic boundary condition (\ref{eqv4}) includes the interface elevation $y(s)$ and its derivatives. Each branch of the interface, $C_-D_-  (c<\xi<\pi)$ and $C_+D_+  (0<\xi<c )$, is evaluated by integration of the derivative of the mapping function (\ref{eqv13}). The parameter form of the interface is as follows,
\begin{equation}
\label{eqv22}
s(\xi)_{\{c<\xi\leq \pi, 0\leq\xi < c\}}=\int_{\{\pi,0\}}^{\xi}\frac{ds}{d\xi}d\xi,
\end{equation}
\begin{equation}
\label{eqv23}
y(\xi)_{\{c<\xi\leq \pi, 0\leq\xi < c\}}=y_D + \Im\left(\int_{\{\pi,0\}}^{\xi}\left.\frac{dz}{d\zeta}\right|_{\zeta=\xi+\pi\tau/4}d\xi\right),
\end{equation}
where
\begin{eqnarray}
\label{eqv24}
\frac{ds}{d\xi}&=&\left|\frac{dz}{d\zeta}\right|_{\zeta=\xi+\pi\tau/4} \nonumber \\
&=&\frac{K}{v(\xi)}\left|\frac{\vartheta_4^2(\xi-a+\pi\tau/4)\vartheta_4^2(\xi-b+\pi\tau/4) }{\vartheta_1^2(\xi-c)\vartheta_4^2(\xi-c)}\right|
\end{eqnarray}
and $y_D$ is the vertical coordinate of points $D_-D_+$ and can be obtained from
\begin{equation}
\label{eqv25}
y_D=\Im\left(i\int_0^{\pi|\tau|/4}\left.\frac{dz}{d\zeta}\right|_{\zeta=i\eta}d\eta\right).
\end{equation}

The curvature of the interface is
\begin{equation}
\label{eqv26}
\kappa[s(\xi)]=\frac{d\beta}{ds}=\frac{d\beta}{d\xi}/\frac{ds}{d\xi},
\end{equation}
where $d\beta/d\xi$ is determined by taking the argument of the complex velocity from equation (\ref{eqv10}) at $\zeta=\xi+\pi\tau/4$,
\begin{eqnarray}
\label{eqv27}
\beta(\xi)&=&\arg\left(\frac{dw}{dz}\right) = -\frac{1}{\pi}\int_\pi^0\frac{d\ln v}{d\xi^\prime}\ln \left|\frac{\vartheta_1(\xi-\xi^\prime)}{\vartheta_4(\xi-\xi^\prime)}\right| - P_1(\xi), \nonumber \\
& &
\end{eqnarray}
\begin{eqnarray}
\label{eqv28}
P_1(\xi)&=&-\beta_O+\Im\left\{\ln\frac{\vartheta_1(\xi-a+\pi\tau/4)\vartheta_1(\xi-b+\pi\tau/4)}{\vartheta_4(\xi-a+\pi\tau/4)\vartheta_4(\xi-b+\pi\tau/4)}\right\} \nonumber \\
&+& \frac{1}{\pi}\int_0^\pi\frac{d\beta_b}{d\xi^\prime}\Im\left\{\ln \frac{\vartheta_1(\xi-\xi^\prime+\pi\tau/4)}{\vartheta_4(\xi-\xi^\prime+\pi\tau/4)}
\right\}d\xi^\prime
\end{eqnarray}
and differentiating it respect to variable $\xi$,
\begin{equation}
\label{eqv29}
\frac{d\beta}{d\xi}=-\frac{1}{\pi}\int_\pi^0 \frac{d\ln v}{d\xi^\prime}\left(\frac{\vartheta_1^\prime(\xi-\xi^\prime)}{\vartheta_1(\xi-\xi^\prime)}-\frac{\vartheta_4^\prime(\xi-\xi^\prime)}{\vartheta_4(\xi-\xi^\prime)}\right)
d\xi^\prime -P_1^\prime(\xi).
\end{equation}
Here, prime denotes the derivatives of the functions with respect to $\xi$. The integrand of the above equation has a first-order singularity at point  $\xi^\prime=\xi$, since $\vartheta_1(\xi-\xi^\prime)\sim \xi-\xi^\prime$.

The derivatives of the curvature, $d\kappa/ds$  and $d^2\kappa/ds^2$, can be obtained by differentiating (\ref{eqv26}). They will include higher-order derivatives of the function $\beta(\xi)$  and a higher-order singularity in the integrands, respectively. By substituting the derivatives of the curvature into the dynamic boundary condition (\ref{eqv4}), we can derive a hypersingular integral equation in terms of the function $d\ln v/d\xi$, the solution of which requires special treatment. Instead of that, we will determine the function $v(\xi)$ numerically using a collocation method.

\subsection{\label{sec26} Numerical method to determine the function $v(\xi)$.}

If the tentative function $v(\xi)$ is given and the system of equations (\ref{eqv14})-(\ref{eqv18}) and the integrodifferential equation (\ref{eqv21}) are solved, then the interface $z=z(\xi)$ depends only on the given function $v(\xi)$. We can chose a fixed set of points $\hat{\xi}_k$, $k=1,\bar{K}$, distributed on the side $D_-D_+$ of the parameter region corresponding to the interface. Then, the nodes $s_k=s(\hat{\xi}_k)$ and the coordinates $y_k=y(\hat{\xi}_k)$ in the physical plane are obtained using Eqs. (\ref{eqv22}) and (\ref{eqv23}). The curvature of the interface and its derivatives are determined numerically applying spline interpolation of the nodes $y_k$  in the intervals $(s_{k-1},s_k)$. We chose a fifth-order spline which provides continuous derivatives along the interface up to the fourth derivative,
\begin{eqnarray}
\label{eqv30}
y(s)=y_k+a_{1,k}(s-s_{k-1})+\ldots+a_{n,k}(s-s_{k-1})^n, \nonumber \\
\quad s_{k-1}<s<s_k,\quad k=1,\ldots,\bar{K}, \quad n=5.
\end{eqnarray}
The curvature and its derivatives are obtained by differentiating Eq. (\ref{eqv30}):
\[
\beta=\arcsin y^\prime, \quad \kappa=\frac{y^{\prime\prime}}{\sqrt{1-y^{\prime 2}}}, \quad  \frac{d\kappa}{ds}=\frac{y^\prime y^{\prime\prime 2}-y^{\prime\prime\prime}(y^{\prime 2}-1)}{(1-y^{\prime 2})^{3/2}}, \cdots.
\]

The system of nonlinear equations can be obtained by applying the dynamic boundary condition (\ref{eqv4}) at the points $s_k=s(\hat{\xi_k})$, which is written in the form
\begin{eqnarray}
\label{eqv31}
G_k(\bar{V})=c_{pk}(\bar{V})-c_{pk}^{ice}(\bar{V})=0, \quad k=1,\ldots, \bar{K},
\end{eqnarray}
where $\bar{V}=(v_1,\ldots,v_{\bar{K}})^{T}$ is the vector of unknown velocities $v_k$ on the interface;
the pressure coefficient on the interface due to the flow is
\begin{eqnarray}
\label{eqv32}
c_{pk}(\bar{V})=1-v_k^2-\frac{2[y(\bar{V})-h]}{F^2};
\end{eqnarray}
and the pressure coefficient determining the bending of the elastic sheet is
\begin{eqnarray}
\label{eqv33}
c_{pk}^{ice}(\bar{V})=2D\left[ \left( \frac{d^2\kappa}{ds^2} \right)_k +\frac{1}{2}\kappa_k^3\right].
\end{eqnarray}

The system of equations (\ref{eqv31}) is solved using Newton's method.The Jacobian of the system is evaluated numerically using the central difference with $\Delta v_k=10^{-5}$, $k=1,\bar{K}$. At each evaluation of the function $G_k(\bar{V})$, the system of equations (\ref{eqv14})-(\ref{eqv18}) and the integrodifferential equation (\ref{eqv21}) are solved. From $5$ to $20$ iterations are necessary to get convergence of the solution. All solutions, say $\bar{V^\ast}$, reported here satisfied the condition
\begin{eqnarray}
\label{eqv34}
\sum_1^{\bar{K}}|G_k(\bar{V^\ast})|<10^{-7}.
\end{eqnarray}
which is regarded as giving a sufficiently accurate solution of the nonlinear equations. However, the inaccuracy within the intervals $(\hat{\xi}_{k-1},\hat{\xi}_k)$ is somewhat lower. It will be discussed in detail in section 3.1.

For supercritical flow regimes, the waving  interface may extend to infinity. However, due to the finite length of the calculation region and the condition that the flow is uninform at infinity in all directions (upstream, downstream and at infinite depth $y\rightarrow-\infty$), we need to introduce the damping regions $P_2P_1$  upstream and $T_1T_2$ downstream. In these regions, we add an artificial term in the boundary condition (\ref{eqv4}) which may be treated as an external applied pressure,
\begin{equation}
\label{eqv35}
c_p=1-v^2-\frac{2(y-h)}{F^2}+C_dv\frac{dv}{ds},
\end{equation}
where the damping coefficient $C_d$ increases from $0$ at points $P_1$ and $T_1$ to the values $C_{dL}$ and $C_{dR}$ at points $P_2$ and $T_2$, respectively. The length of the damping regions are chosen to be about $2\lambda_0$, where $\lambda_0$ is the wave length of the free surface progressive waves according to the first approximation theory \cite{Lamb}.

\subsection{\label{sec27} Dispersion equation.}

Differentiating Eq. (\ref{eqv4}) with respect to the arc length coordinate along the interface and taking into account that the slope of the interface $\delta=\arcsin(dy/ds)$ is the angle between the unit tangential vector $\mathbf{\tau}$ and the $x-$axis, we obtain
\begin{equation}
\label{eqv36}
F^2v^2\frac{d\ln v}{ds}=-\sin\delta-F^2 D\left[ \frac{d^4\beta}{ds^4}+\frac{3}{2}\left(\frac{d\beta}{ds}^2 \right)\frac{d^2\beta}{ds^2}\right].
\end{equation}
The above equation without an elastic sheet ($D=0$) and at small disturbances of the free surface such that $\sin\delta\approx\delta$ can be written as \begin{equation}
\label{eqv37}
v^2\frac{d\ln v}{ds}=-\frac{2\pi}{\lambda}\delta
\end{equation}
where the wave length $\lambda=\lambda_0=2\pi F^2$  is known from the linear theory of gravity waves \cite{Lamb}.
In the presence of the sheet ($D\neq0$), the waves of small amplitude approach a sinusoidal curve. Therefore, their slope can be written as
\[
\delta(s)=\delta_{max} \cos\left(\frac{2\pi s}{\lambda} + \phi\right),
\]
where $\delta_{max}$ is the amplitude, and $\phi$ is the phase of the slope relative to point $D_+$ at which $s=0$. Then, neglecting the square of the curvature, i.e. the second term in brackets (\ref{eqv36}), we obtain
\begin{equation}
\label{eqv38}
v^2\frac{d\ln v}{ds}=-\delta\left[\frac{1}{F^2}+D\left(\frac{2\pi}{\lambda}\right)^4 \right].
\end{equation}

According to (\ref{eqv37}), the coefficient at the angle $\delta$  is equal to $-2\pi/\lambda$, where $\lambda$  is the wave length of the interface. Therefore, the following equation with respect to wave number $k=2\pi/\lambda$ is obtained	
\begin{equation}
\label{eqv39}
k=\frac{1}{F^2}+Dk^4.
\end{equation}
This equation may have one, two or no roots, which depends on the constant $D$ depending on the thickness of the elastic sheet and the Froude number $F$. The latter case corresponds to the subcritical flow regime for which the perturbation on the interface decays in both directions. The case of one root corresponds to the critical Froude number, $F_{cr}$, for which the waves of the same length $\lambda_{cr}=2\pi/k_{cr}$  are extended to infinity in both directions. Differentiating (\ref{eqv39}) with respect to $k$ and equating the result to zero, after some manipulations we obtain
\begin{equation}
\label{eqv40}
k_{cr}=\sqrt[3]{4}\left(\frac{\rho g L (1-\nu^2)}{EB_i^3} \right)^{\frac{1}{4}}, \quad   F_{cr}=\left(\frac{64}{81}\frac{EB_i}{\rho g L (1-\nu^2)} \right)^{\frac{1}{8}}.
\end{equation}

For $F>F_{cr}$, Eq. (\ref{eqv39}) has two roots, $k_w<k_{cr}$  due to gravity and $k_{ice}>k_{cr}$  due to the elastic sheet. We note that the Eq. (\ref{eqv38}) is valid along the whole interface, $-\infty<x<\infty$, so both waves associated with wave numbers $k_w$  and $k_{ice}$ may appear upstream and downstream of the submerged cylinder. We note that the depth of submergence of the body does not appear in the dispersion equation, and therefore it does not influence the wave number. However, we expect that the depth of submergence influences the wave amplitude, similar to that observed for free-surface flows \cite{Sem_Wu2020}.

\begin{figure}
\centering
\includegraphics[scale=0.38]{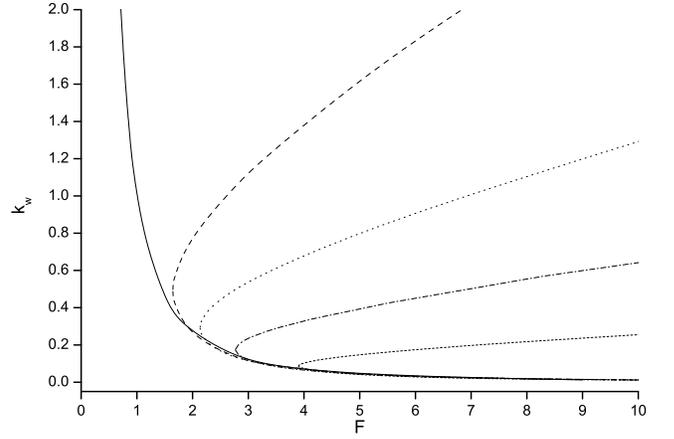}
\vspace*{-20mm}
\caption{Wave number vs. Froude number for different thicknesses of the elastic: $b_i=0$ (solid line), $0.05$ (dashed line), $0.1$ (dotted line), $0.2$ (dot-dashed line), $0.5$ (short dashed line).}
\label{figure2}
\end{figure}

The roots of equation (\ref{eqv39}) for different Froude numbers and different thicknesses of the elastic sheet are shown in figure \ref{figure2}.

Without an elastic sheet, the constant $D=0$, and Eq. (40) becomes $k=1/F^2$ that corresponds to the hyperbola in figure $\ref{figure2}$. For thickness $b_i >0$, the critical Froude number appears as the minimal Froude to which corresponds only one root and the vertical orientation of the slope. The larger relative thickness of the elastic sheet results in the larger critical Froude number. The dispersion equation predicts only possible waves, but the contribution of each wave to the shape of the interface have to be determined from the solution of the nonlinear problem.

\section{\label{sec3} Results}
\subsection{\label{sec31} Numerical approach.}
In the discrete form, the solution is sought on a fixed set of points $\xi_j$, $j=1,N$, distributed along the side $O_-O_+$, $0\leq \xi \leq \pi$, $\eta=0$, and a fixed set of points $\hat{\xi}_i$, $i=1,M$ distributed along the side $D_-D_+$, $\eta=\pi/4$, of the parameter region. The points $\xi_j$ are distributed so as to provide a higher density of the points $s_j=s_b(\xi_j)$ near stagnation points $A (\zeta=a)$ and $B (\zeta=b)$. The distribution of the points $\hat{\xi}_i$ is chosen such to provide a higher density of the points $s_i=s(\hat{\xi}_i)$ closer to the body and the uniform distribution for $|s_i|>\lambda$.

The number of nodes on the body and the interface are chosen in the ranges $N=100\div200$  and $M=1000\div5000$, respectively, based on the requirement to provide at least $80$ nodes within the shorter waves to get convergence and reasonable accuracy of the solution. The nodes $\hat{\xi}_k$, $k=1,\bar{K}$, used for interpolation of the interface, $y_k=y(\hat{\xi}_k)$, $s_k=s(\hat{\xi}_k)$, are chosen on the set of points $\hat{\xi}_i$ such that $i=4k$. Then, $\bar{K}=M/4=250\div1000$ provides 20 nodes within the shorter wave length at which the system of the nonlinear equations (\ref{eqv31}) is solved.

The integrals appearing in Eq. (\ref{eqv10}) are evaluated based on the linear interpolation of the functions $\beta_b(\xi)$ and $\ln v(\xi)$ on the segments $(\xi_{j-1},\xi_j)$ and $(\hat{\xi}_{i-1},\hat{\xi}_i)$, respectively. They are evaluated as follows:
\begin{eqnarray}
\label{eqv41}
\frac{1}{\pi}\int_0^\pi \frac{d\beta_b}{d\xi}\ln\left(\frac{\vartheta_1(\zeta-\xi)}{\vartheta_4(\zeta-\xi)}\right)d\xi=\Delta\beta_{bj}[r_j(\zeta)+iq_j(\zeta)], \nonumber \\ j=1,\ldots,N.     \quad
\end{eqnarray}
\begin{eqnarray}
\label{eqv42}
\frac{i}{\pi}\int_\pi^0 \frac{d\ln v}{d\xi}\ln\left(\frac{\vartheta_1(\zeta-\xi-\pi\tau/4)}{\vartheta_4(\zeta-\xi-\pi\tau/4)}\right)d\xi=\Delta\ln v_i[\hat{r}_i(\zeta)+i\hat{q}_i(\zeta)], \nonumber \\ i=1,\ldots,M.   \qquad
\end{eqnarray}
where
$\Delta\beta_{bj}=\beta_b(\xi_j)-\beta_b(\xi_{j-1})$, $\Delta\ln v_i =\ln v(\hat{\xi}_i)-\ln v(\hat{\xi}_{i-1})$,
\begin{eqnarray}
\label{eqv43}
r_j(\zeta)=\frac{1}{\pi\Delta\xi_j}\int_{\xi_{j-1}}^{\xi_j}\ln\left|\frac{\vartheta_1(\zeta-\xi)}{\vartheta_4(\zeta-\xi)}\right|d\xi,
\end{eqnarray}
\begin{eqnarray}
\label{eqv44}
q_j(\zeta)=\frac{1}{\pi\Delta\xi_j}\int_{\xi_{j-1}}^{\xi_j}\arg\left(\frac{\vartheta_1(\zeta-\xi)}{\vartheta_4(\zeta-\xi)}\right)d\xi,
\end{eqnarray}
\begin{eqnarray}
\label{eqv45}
\hat{r}_i(\zeta)=-\frac{1}{\pi\Delta\hat{\xi}_i}\int_{\hat{\xi}_{i-1}}^{\hat{\xi}_i}\arg\left(\frac{\vartheta_1(\zeta-\xi-\pi\tau/4)}{\vartheta_4(\zeta-\xi-\pi\tau/4)}\right)d\xi,
\end{eqnarray}
\begin{eqnarray}
\label{eqv46}
\hat{q}_i(\zeta)=\frac{1}{\pi\Delta\hat{\xi}_i}\int_{\hat{\xi}_{i-1}}^{\hat{\xi}_i}\ln\left|\frac{\vartheta_1(\zeta-\xi-\pi\tau/4)}{\vartheta_4(\zeta-\xi-\pi\tau/4)}\right|d\xi,
\end{eqnarray}
The integrals (\ref{eqv43}) - (\ref{eqv46}) are evaluated using the $8-$point Legendre-Gauss quadrature formula. The error of the solution within the intervals of interpolation $(\hat{\xi}_{k-1},\hat{\xi}_{k})$ satisfies the relation
\begin{eqnarray}
\label{eqv47}
\sum_{4k-4}^{4k}|G_i(\bar{V}^\ast)|< 10^{-3}
\end{eqnarray}
which is regarded as giving a sufficiently accurate solution of the problem.

The method of successive approximations is adopted to solve the integrodifferential equation (\ref{eqv21}), which in the discrete form becomes
\begin{equation}
\label{eqv48}
\frac{(\Delta \beta_b)^{(k+1)}_j}{\Delta \xi_j}=\frac{\beta_b[s_b^{(k)}(\xi_j)] - \beta_b[s_b^{(k)}(\xi_{j-1})]}{\Delta \xi_j}, \qquad j=1,\ldots,N,
\end{equation}
where  the arc length along the body, $s_b^{(k)}(\xi)$, is evaluated using (\ref{eqv19}) with $(\Delta \beta_{bj}/\Delta \xi_j)^{(k)}$ known at the $(k)^{th}$ iteration. The iteration process converges very fast. After $5$ to $10$ iterations the error is below a prescribed tolerance of $10^{-6}$.

The derivative of the mapping function (\ref{eqv13}) has a second-order singularity at point $\zeta=c+\pi\tau/4$. Therefore, points $\hat{\xi_i}$, $i=1,M$,  along the side $D_-D_+$ in the parameter region are distributed within the two intervals $c+\varepsilon_1<\hat{\xi}_i\leq\pi$, $i=1,M_1$, and $0\leq \hat{\xi}_i<c-\varepsilon_2$, $i=M_1+1,M$. These intervals correspond to parts $D_-C_-$ and $D_+C_+$ of the interface $C_-C_+$ in the physical plane. The values $\varepsilon_1$  and $\varepsilon_2$  are chosen to provide the required length of the parts $D_-C_-$  and $D_+C_+$.

\subsection{\label{sec32} Convergence study of the numerical method.}

\begin{figure}
\centering
\includegraphics[scale=0.58]{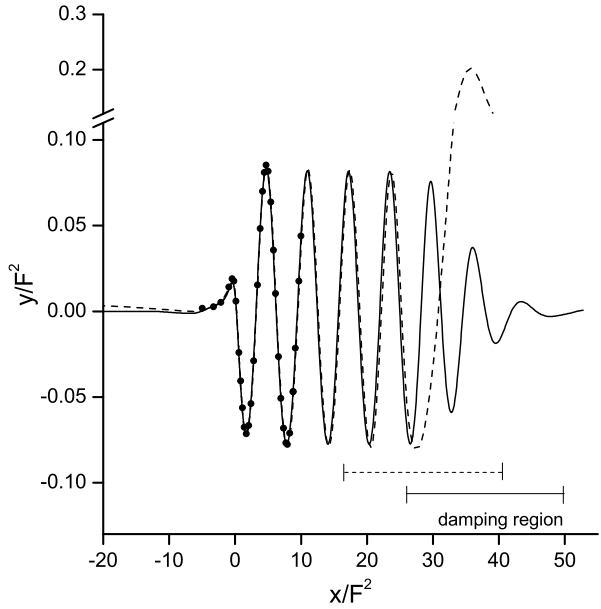}
\vspace*{-20mm}
\caption{Verification of the numerical procedure comparing the shape of the free surface using different methods: present collocation method (solid line); successive approximation \cite{Sem_Wu2020} (dashed line); numerical solution \cite{Scullen} (symbols). Froude number $F=2.75$, depth of submergence $h=7.55$.}
\label{figure3}
\end{figure}

The formulation of the problem allows us to consider the free-surface flow around the submerged circular cylinder if we chose zero thickness of the elastic sheet. This case has been investigated in \cite{Sem_Wu2020} using the method of successive approximations. The results based on the present collocation method and that based on the successive approximations are shown in Fig. \ref{figure3}.
Without elastic sheet, the submerged body generates a progressive wave downstream only. The free surface upstream for $x<-\lambda$ tends to be parallel to the $x-$axis. This property was used in the model of Semenov \& Wu \cite{Sem_Wu2020} to determine a parameter which affects the velocity magnitude   in the damping region. Both the present numerical method and the computational procedures\cite{Sem_Wu2020} predict the same shape of the free surface in the region $s_{P1}<s<s_{T1}$. However, the present damping model provides the gradual decay of the wave that is necessary to couple the pressure coefficients due to the flow and the bending of the elastic sheet in the numerical procedure. It is seen that the truncation of the computational region does not affect the shape of the free surface in the interval $s_{P1}<s<s_{T1}$, as seen in Fig. \ref{figure3}.

\begin{figure}
\centering
\includegraphics[scale=0.42]{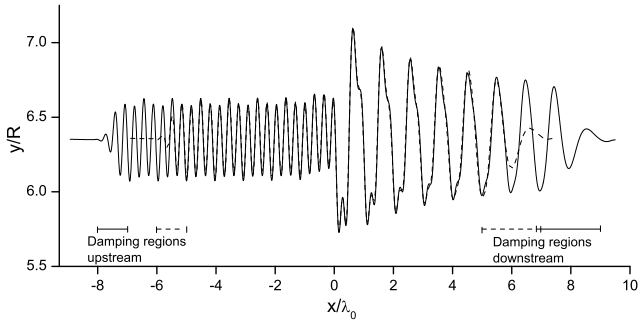}
\vspace*{-15mm}
\caption{Flexural-gravity waves generated by the submerged circular cylinder. Thickness of elastic sheet $b_i=0.05$, Froude number $F=2$, critical Froude number $F_{cr}=1.65$, depth of submergence $h=6.35$.}
\label{figure4}
\end{figure}
In order to investigate the effect of truncation in the presence of the sheet, we consider an example of supercritical flow for the thickness $b_i=0.05$  and the depth of submergence $6.35$. From Eq. (\ref{eqv39}), we obtain the critical Froude number, $F_{cr}=1.65$. For Froude number $F=2$, the flow is supercritical, so there are two wave numbers $k_w=0.256$ and $k_{ice}=0.782$ determined from the dispersion equation (\ref{eqv39}). These wave numbers correspond to the wave lengths $\lambda_w=24.5R$ and $\lambda_{ice}=8.03R$. The wave length corresponding to the linear theory without an elastic sheet is $\lambda_0=2\pi F^2R$. The interface is shown in Fig. \ref{figure4} for two computational regions $-8<x/\lambda_0<9$  (solid line) and $-6<x/\lambda_0<7$ (dashed line). In the region $-5<x/\lambda_0<5$, where the damping is absent ($C_d=0$) for both cases, the interfaces overlap. Outside this region, the solid lines and the dashed lines start to diverge. These results show that similar to free-surface flows, the truncation of the computational region does not affect the part of the interface without damping. The computations in Fig. \ref{figure4} are carried out for the length of the damping region, $x_{P1} - x_{P2}=\lambda_0$, and the damping region downstream, $x_{T2} - x_{T1}=2\lambda_0$. The values $C_{dL}=2$  and $C_{dR}=10$  were used, and the numbers of nodes are $N=100$ and $M=4000$.

\subsection{\label{sec32} Subcritical flows.}
\begin{figure}
\centering
\includegraphics[scale=0.5]{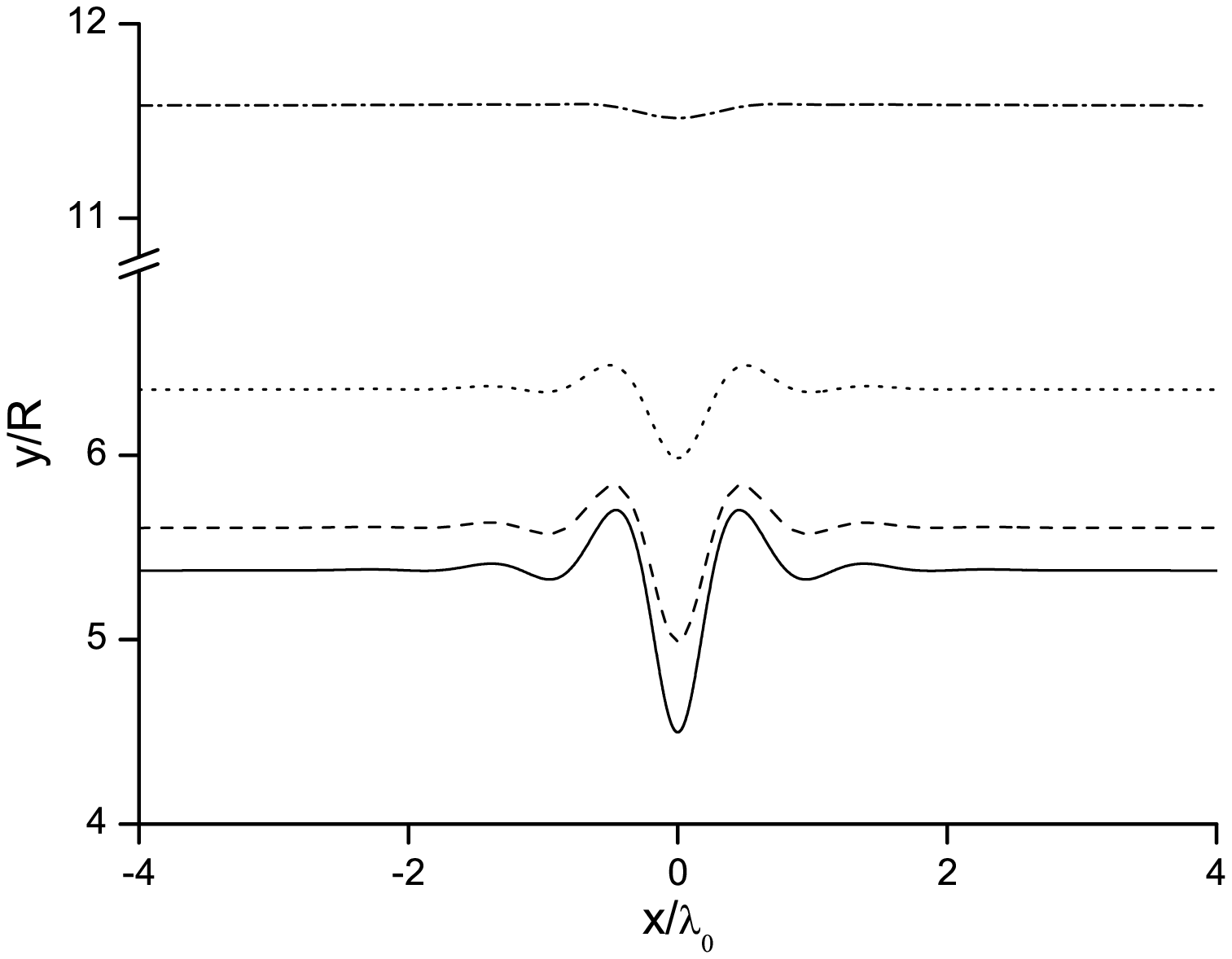}\\
\vspace*{-15mm}
($a$)
\vspace*{-15mm}
\includegraphics[scale=0.5]{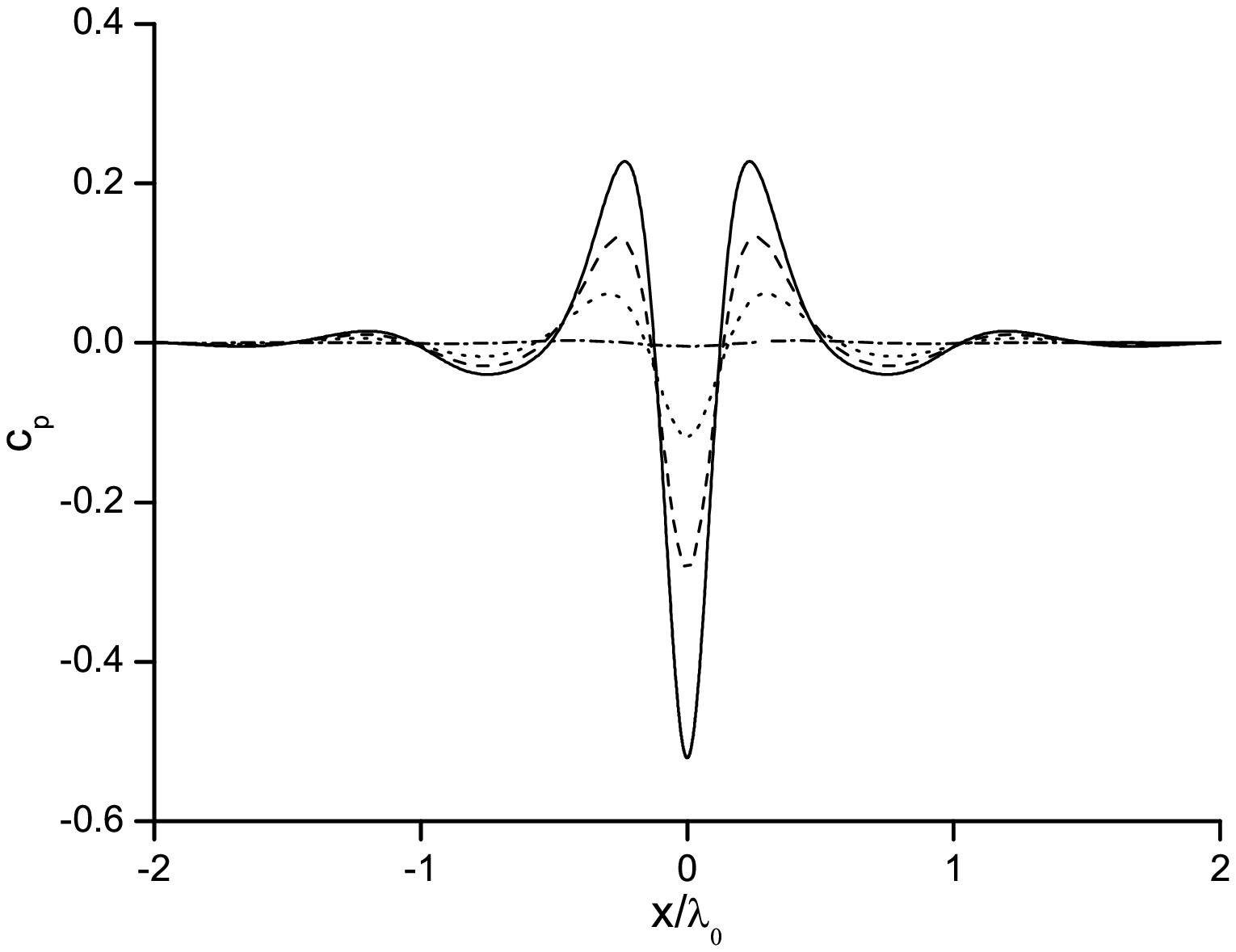}\\
($b$)
\caption{Free surface shape (a) and pressure coefficient (b) for Froude number $F=1.5$, $F_{cr}=1.65$ and depth of submergence $h=5.37$ (solid line), $5.60$ (dashed line), $6.35$ (dotted line) and $11.58$ (dash-dotted line). }
\label{figure5}
\end{figure}

For Froude numbers $F<F_{cr}$, Eq. (\ref{eqv39}) has only complex roots, which correspond to decaying perturbations of the interface caused by the submerged cylinder. In Fig. \ref{figure5}, we show the interface profiles for the Froude number $F=1.5$ and the relative thickness of the elastic sheet $b_i=0.05$ for different depths of submergence. The interface shape is symmetric about the $y-$axis. The shape is different from that observed for the free-surface flow without an elastic sheet, for which the free surface is almost flat upstream and exhibits a wave downstream. Thus, the elastic sheet supresses the waves downstream and perturbs the flow upstream near the cylinder. As the thickness of the elastic sheet tends to zero, the critical Froude number $F_{cr}$ decreases and become smaller than $F$. In this case, the flow becomes supercritical, which drastically changes the interface shape. It will be studied in the following subsection.

It is expected that if cylinder is closer to the elastic sheet, or the depth of submergence is smaller, then interaction between the cylinder and the elastic sheet is stronger. This is observed in Fig. \ref{figure5}. The deflection of the sheet above the cylinder exhibits a trough making the gap between the sheet and the cylinder smaller. It is found that there is a minimal, or critical depth of submergence, $h_{cr}$, below which the numerical solution cannot be obtained. For $h$ slightly larger than $h_{cr}$, few iterations to solve the system of nonlinear equations (\ref{eqv30}) are required to get the converged  solution, while for $h$ slightly smaller than $h_{cr}$, the elastic sheet starts to oscillate, and the iterations do not converge.

\begin{figure}
\centering
\includegraphics[scale=0.5]{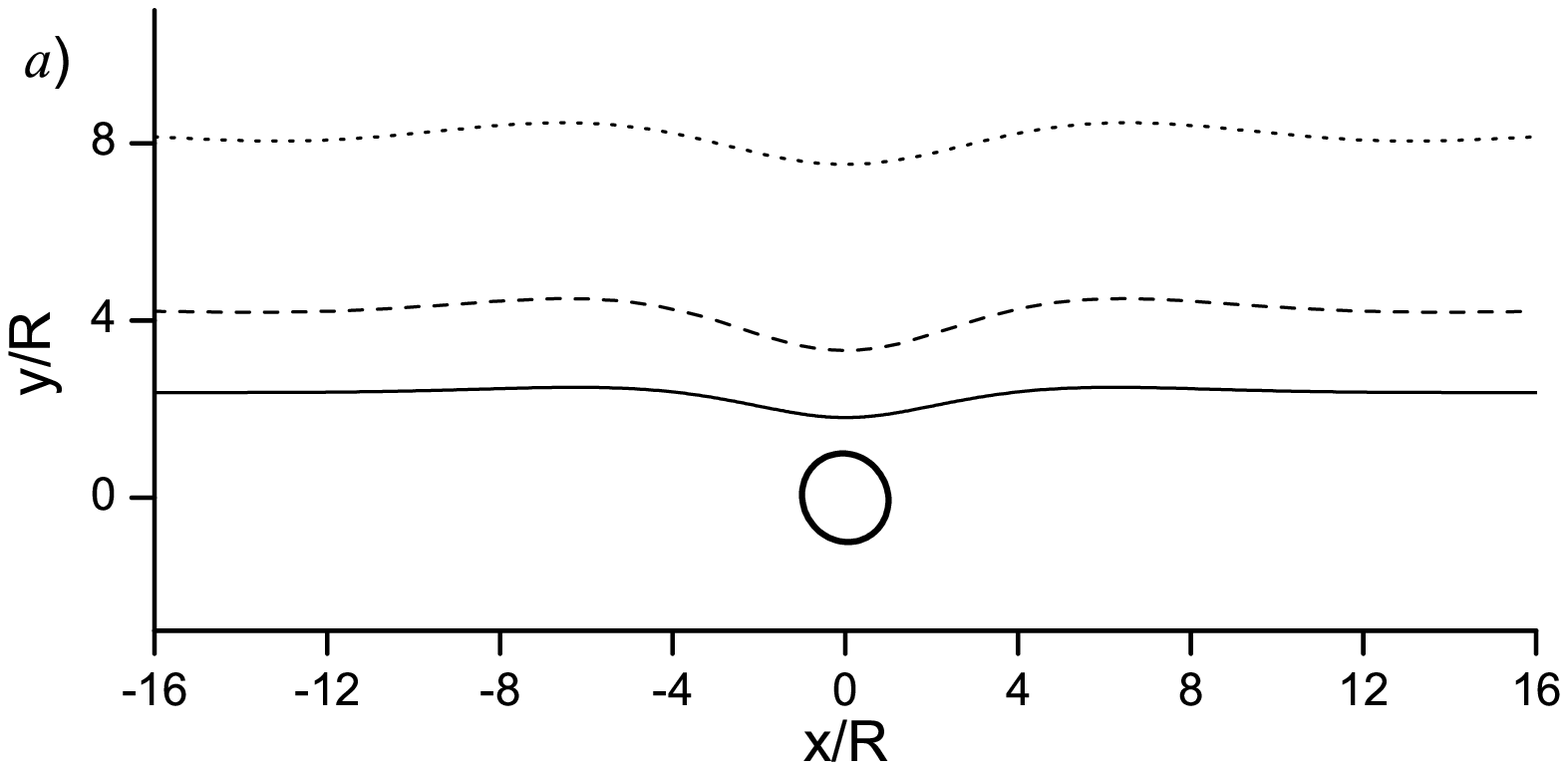}\\
\vspace*{-10mm}
\includegraphics[scale=0.5]{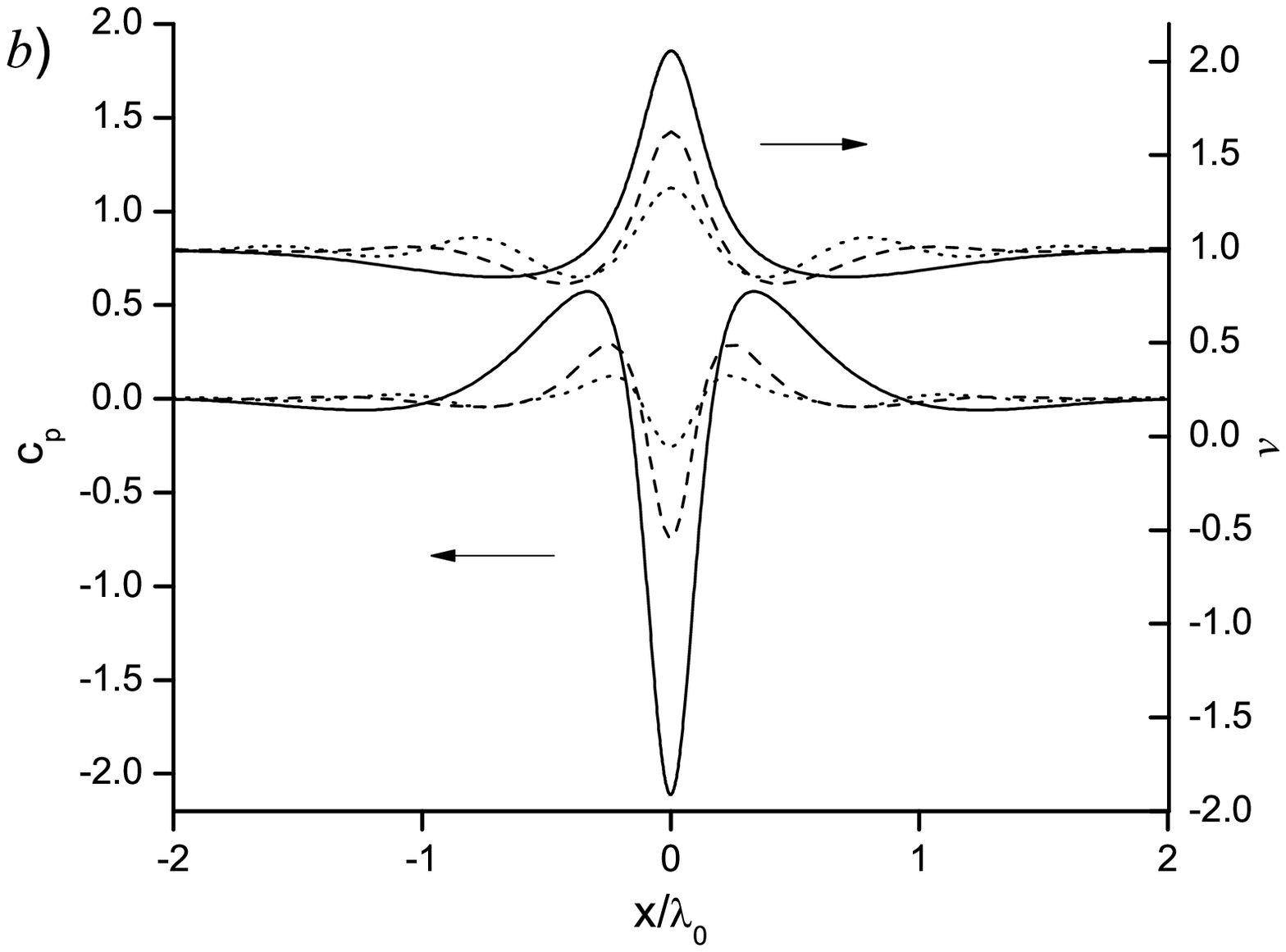}\\
\vspace*{-20mm}
\caption{Free surface shape (a) and pressure coefficient (b) for Froude number $F=1.5$, $F_{cr}=1.65$ and depth of submergence $h=5.37$ (solid line), $5.60$ (dashed line), $6.35$ (dotted line) and $11.58$ (dash-dotted line). }
\label{figure6}
\end{figure}
The interface profiles corresponding to   for different Froude numbers are shown in Fig. 6a. It is seen that the critical depth becomes larger as the Froude number approaches the critical value  . The corresponding distributions of the pressure coefficient (the left axis) and the velocity magnitude (the right axis) along the interface are shown in Fig. 6b. For the case of free-surface flows, there is also a depth of submergence below which a steady free-surface flow does not exist. In that case, the velocity magnitude at the crest of the waves tends to zero, and the free surface shape forms a corner of  . The mechanism restricting the existence of the steady flow in the presence of the elastic sheet is different. As shown in Fig. 6a, the velocity magnitude on the interface is much larger than zero. The present results show that there is a condition of consistency in interaction between the fluid, elastic sheet and the submerged cylinder.

\subsection{\label{sec32} Supercritical flows.}
We begin the computational analysis for relatively high Froude number, $F=3$, or $F/F_{cr}=1.82$, and the ice thickness $b_i=0.05$, for which the dispersion equation (\ref{eqv39}) has two real roots. The corresponding wave numbers are $k_{ice}=1.13$ and $k_w=0.1113$. The second wave number almost coincides with that obtained from the linear theory of gravity waves without an elastic sheet,  $k_0=1/F^2=0.1111$. The ratio of the wave lengths generated by the submerged cylinder and the elastic sheet is $\lambda_{ice}/\lambda_w=10.11$.

\begin{figure}
\centering
\includegraphics[scale=0.4]{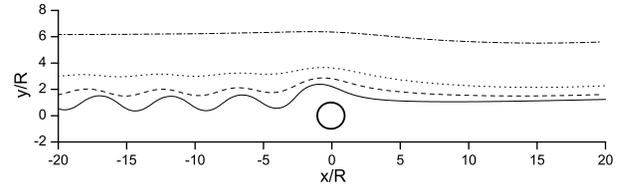}
\vspace*{-10mm}
\caption{Interface shape at different depths of submergence for Froude number $F=3$ and thickness of elastic sheet $b_i=0.05$.}
\label{figure7}
\end{figure}

\begin{figure}
\centering
\includegraphics[scale=0.52]{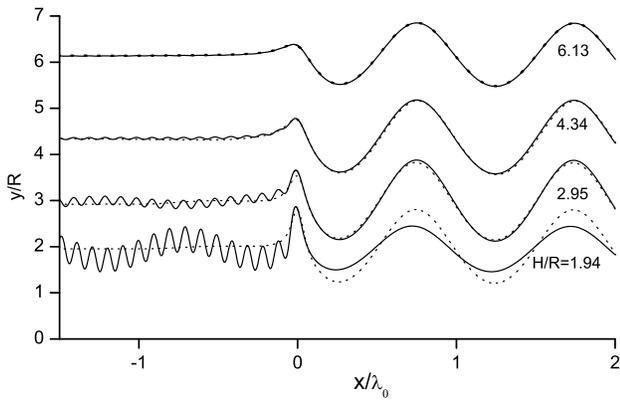}\\
\vspace*{-15mm}
\caption{Effect of submergence on the interface shape for the circular cylinder at Froude number $F=3$. The thickness of elastic sheet $b_i=0.05$ (solid lines) and $b_i=0$ (dashed lines).}
\label{figure8}
\end{figure}

The interfaces near the cylinder are shown in Fig. \ref{figure7} for different depths of submergence. The wave generated by the elastic sheet clearly seen upstream at the smaller depths, and its amplitude decays as the depth of submergence increases. For $h=6.13$ the wave upstream almost disappears. The wave generated by the cylinder downstream is not completely seen since its length   exceeds the length of the interface shown in Fig. \ref{figure7}. The larger length of the interface is shown in Fig. \ref{figure8}. For submergence $h=6.13$, the interface coincides with that obtained without an elastic sheet which is shown in Fig. \ref{figure8} by the dashed line.
As the submergence decreases, the amplitude of the wave downstream increases and reaches its maximal value at $h=2.95$, and then it starts to decrease. This feature has been studied in \cite{Sem_Wu2020}. As the cylinder approaches the free surface, it affects the free surface at smaller distances from the cylinder, and the flow tends to be symmetric about the $y-$axis, similar to that for $F\rightarrow \infty$.

The elastic sheet weakly influences the interface downstream but generates wave upstream. As the cylinder approaches the free surface, the bending of the elastic sheet near the cylinder increases, as can be seen from Figs. \ref{figure7} and \ref{figure8}. This causes the increase of the amplitude of the wave upstream.

\begin{figure}
\centering
\includegraphics[scale=0.52]{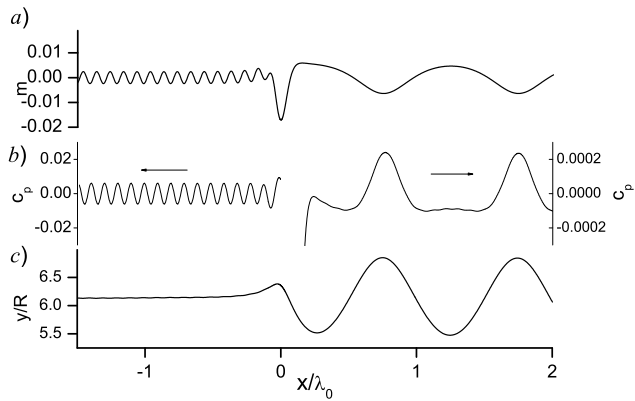}\\
\vspace*{-15mm}
\caption{Bending moment ($a$), pressure coefficient ($b$) and interface shape ($c$) at Froude number $F=3.0$. The thickness of elastic sheet $b_i=0.05$  and depth of submergence $h=6.13$.}
\label{figure9}
\end{figure}

\begin{figure}
\centering
\includegraphics[scale=0.52]{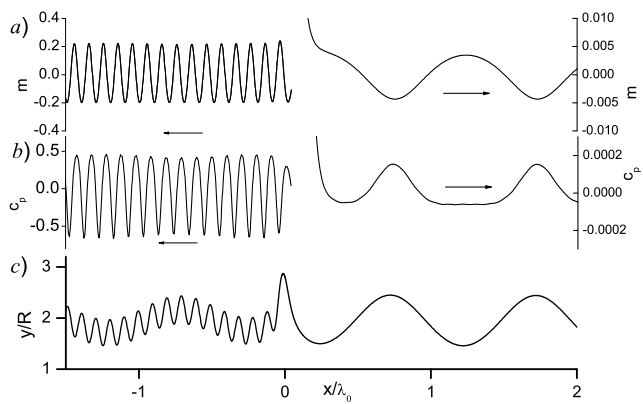}\\
\vspace*{-15mm}
\caption{The same as in Fig. 9 at the depth of submergence $h=1.94$.}
\label{figure10}
\end{figure}

The bending moment and the pressure coefficient are shown in Figs. \ref{figure9} and \ref{figure10} for depths of submergence   and  , respectively. Although the wave due to the elastic sheet is invisible (Fig. \ref{figure9}$c$), the pressure coefficient and the bending moment oscillate at both directions upstream (left axis in Fig. \ref{figure9}$b$) and downstream of the cylinder (right axis in Fig. \ref{figure9}$b$). The frequency of oscillations at the upstream is caused by the elastic sheet, while the frequency of oscillations at the downstream is due to gravity. This qualitatively agrees with results based on the linear theory \cite{Savin_2012,Li_2019}.

\begin{figure}
\centering
\includegraphics[scale=0.4]{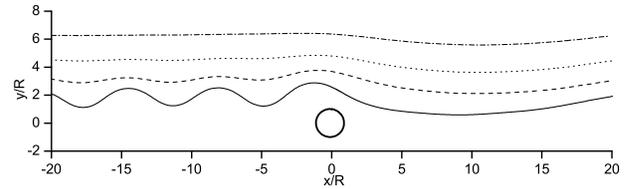}
\vspace*{-10mm}
\caption{Interface shape at different depths of submergence for Froude number $F=2.5$ and thickness of elastic sheet $b_i=0.05$.}
\label{figure11}
\end{figure}

\begin{figure}
\centering
\includegraphics[scale=0.52]{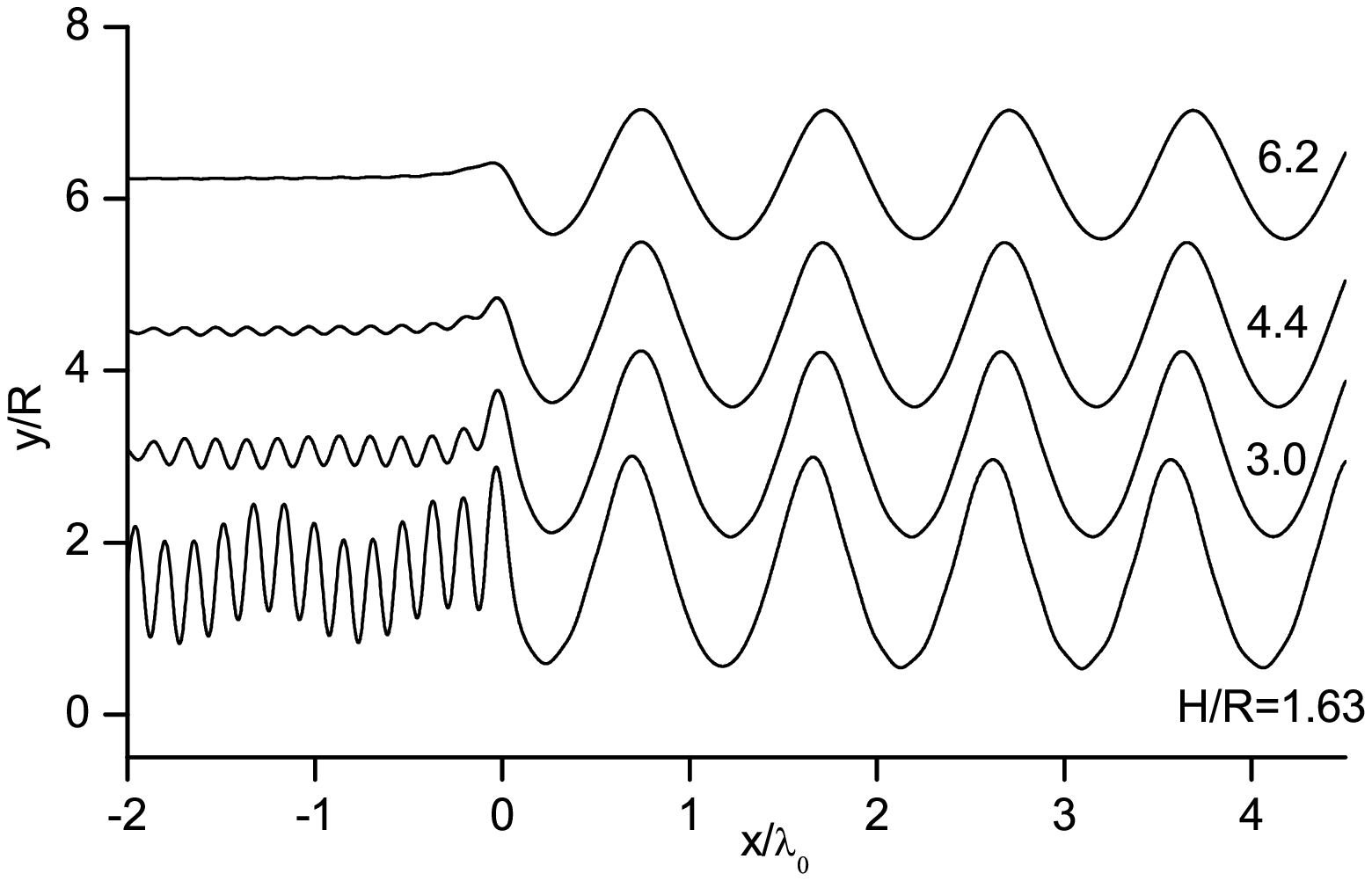}\\
\vspace*{-15mm}
\caption{Effect of submergence on the interface shape for the circular cylinder at Froude number $F=2.5$.}
\label{figure12}
\end{figure}

\begin{figure}
\centering
\includegraphics[scale=0.52]{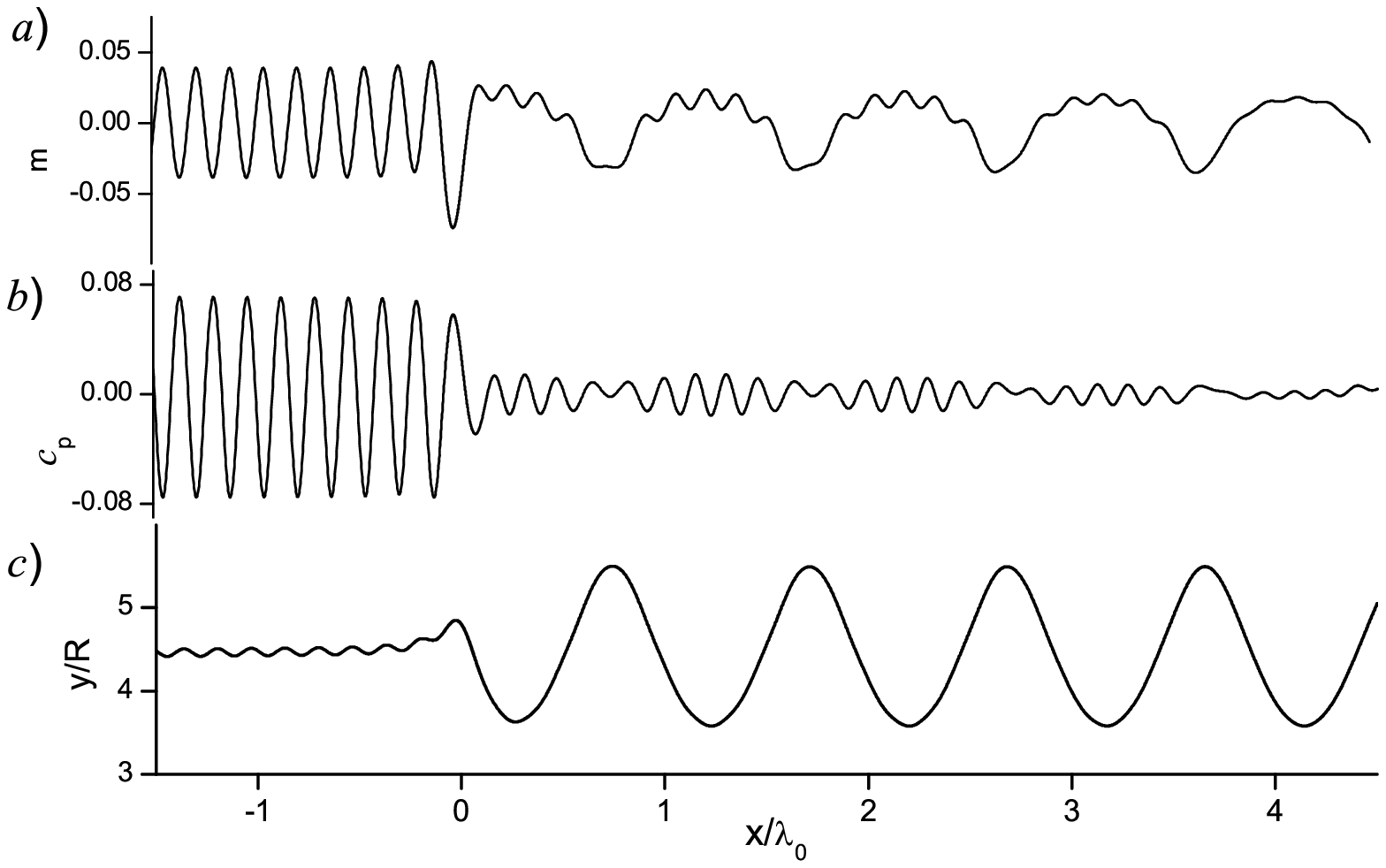}\\
\vspace*{-15mm}
\caption{Bending moment ($a$), pressure coefficient ($b$) and interface shape ($c$) at Froude number $F=2.5$ and the submergence $h=4.4$.}
\label{figure13}
\end{figure}

\begin{figure}
\centering
\includegraphics[scale=0.52]{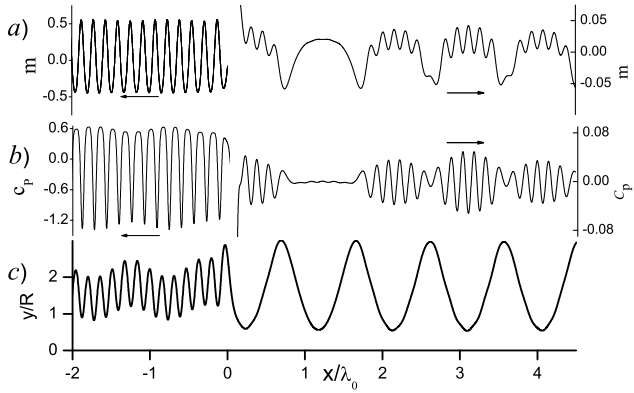}\\
\vspace*{-15mm}
\caption{The same as in Fig. 13 at the depth of submergence $h=1.94$.}
\label{figure14}
\end{figure}

The results for Froude number $F=2.5$, or $F/F_{cr}=1.52$, and ice thickness $b_i=0.05$ are shown in Figs. \ref{figure11}-\ref{figure14}. The wave numbers are as follows: $k_{ice}=0.97$, $k_w=0.1606$ and $k_0=0.1600$. The ratio of the wave lengths $\lambda_w/\lambda_{ice}=6.04$. The interface shapes for different depths of submergence are shown in Fig. \ref{figure11} and \ref{figure12}. They are similar to those in Figs. \ref{figure7} and \ref{figure8} for $F = 3.0$. However, the pressure coefficient and the bending moment shown in Fig. \ref{figure13}$a$,$b$ exhibit behaviour corresponding to superposition of the gravity wave (longer wave) and the elastic sheet wave (shorter wave). The amplitude of the bending moment corresponding to the gravity wave is larger than that corresponding to the elastic sheet wave. The latter becomes largest at the trough of the gravity wave, and it almost disappears at the crest, as seen in Figs. \ref{figure13}$b$ and \ref{figure14}$b$. Such behaviour of the bending moment and the pressure coefficient demonstrates the nonlinear interaction of the elastic sheet and the flow, which is still invisible for the interface profile in Figs. \ref{figure13}$c$ and \ref{figure14}$c$.

\begin{figure}
\centering
\includegraphics[scale=0.4]{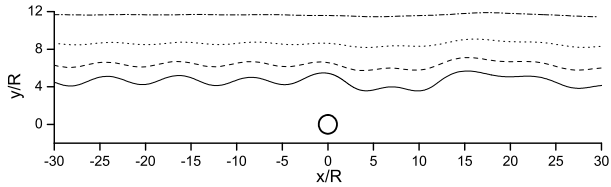}
\vspace*{-10mm}
\caption{Interface shape at different depths of submergence for Froude number $F=2.0$.}
\label{figure15}
\end{figure}

\begin{figure}
\centering
\includegraphics[scale=0.52]{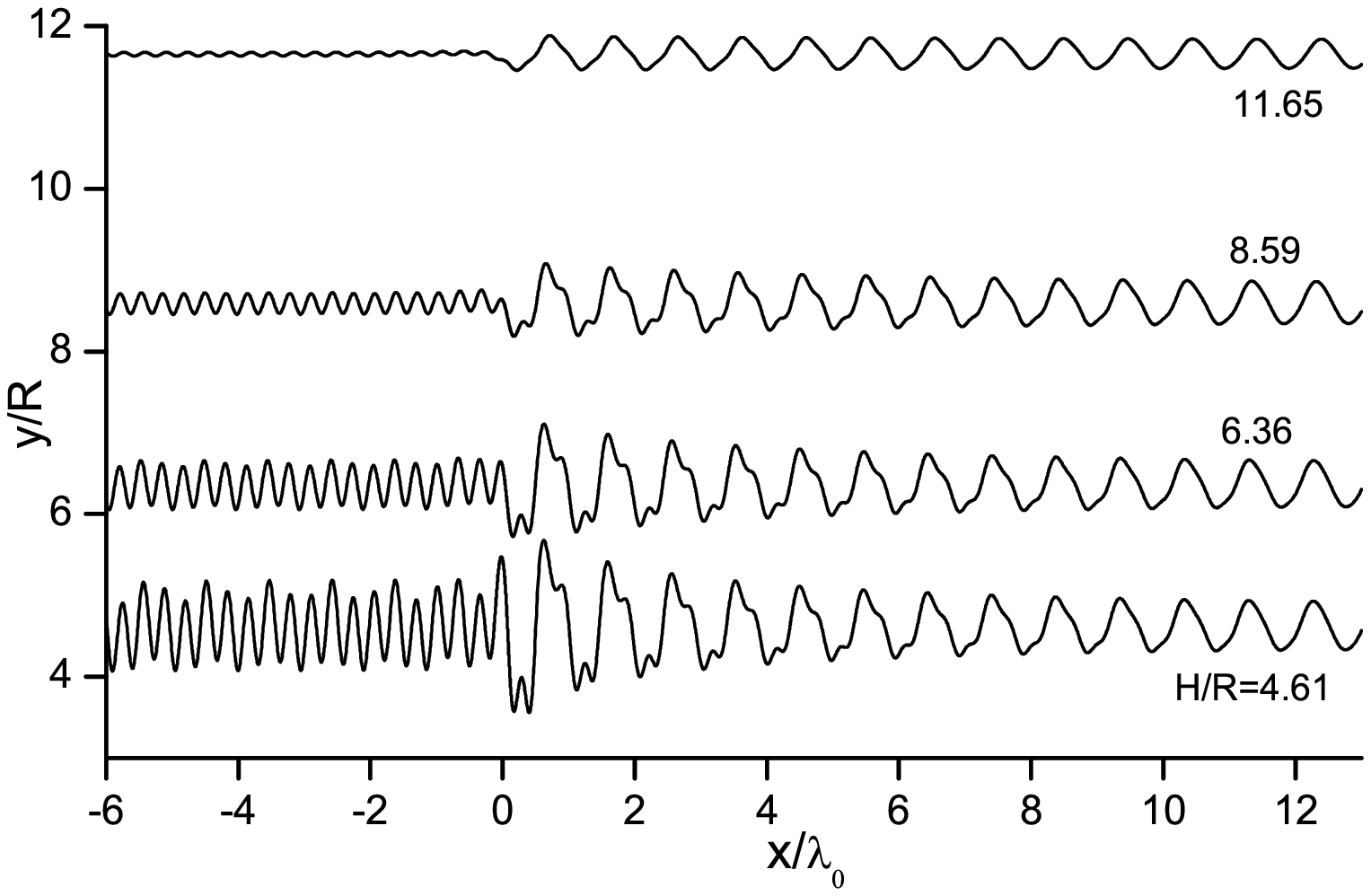}\\
\vspace*{-15mm}
\caption{Effect of submergence on the interface shape for the circular cylinder at Froude number $F=2.0$.}
\label{figure16}
\end{figure}

For Froude number $F=2$, or $F/F_{cr}=1.21$, the wave numbers are as follows: $k_{ice}=0.782$, $k_w=0.256$ and $k_0=0.250$. The ratio of the wave lengths $\lambda_w/\lambda_{ice}=3.05$. The interface shapes for depths of submergence in the range from $4.6$ to $11.6$ are shown in Fig. \ref{figure15} and \ref{figure16}.

At the upstream direction, the wave caused by the elastic sheet becomes visible even for the relatively large depth of submergence, $h=11.6$. Its shape is like a sinusoid with constant amplitude. The amplitude increases as the depth of submergence decreases. At depth of submergence $h=4.61$, the elastic sheet interacts with the flow in such a way that the wave due to gravity extends in the upstream direction, and the interface exhibits superposition of the both waves.

At the downstream direction, the interface shape differs from a wave of constant amplitude. The shape near the body corresponds to the superposition of the gravity and elastic sheet waves, and then gradually approaches to the pure gravity wave. This also occurs for larger submergence but is less visible. The contribution of the elastic sheet wave decays downstream, and the interface approaches the pure gravity wave far downstream.
\begin{figure}
\centering
\includegraphics[scale=0.52]{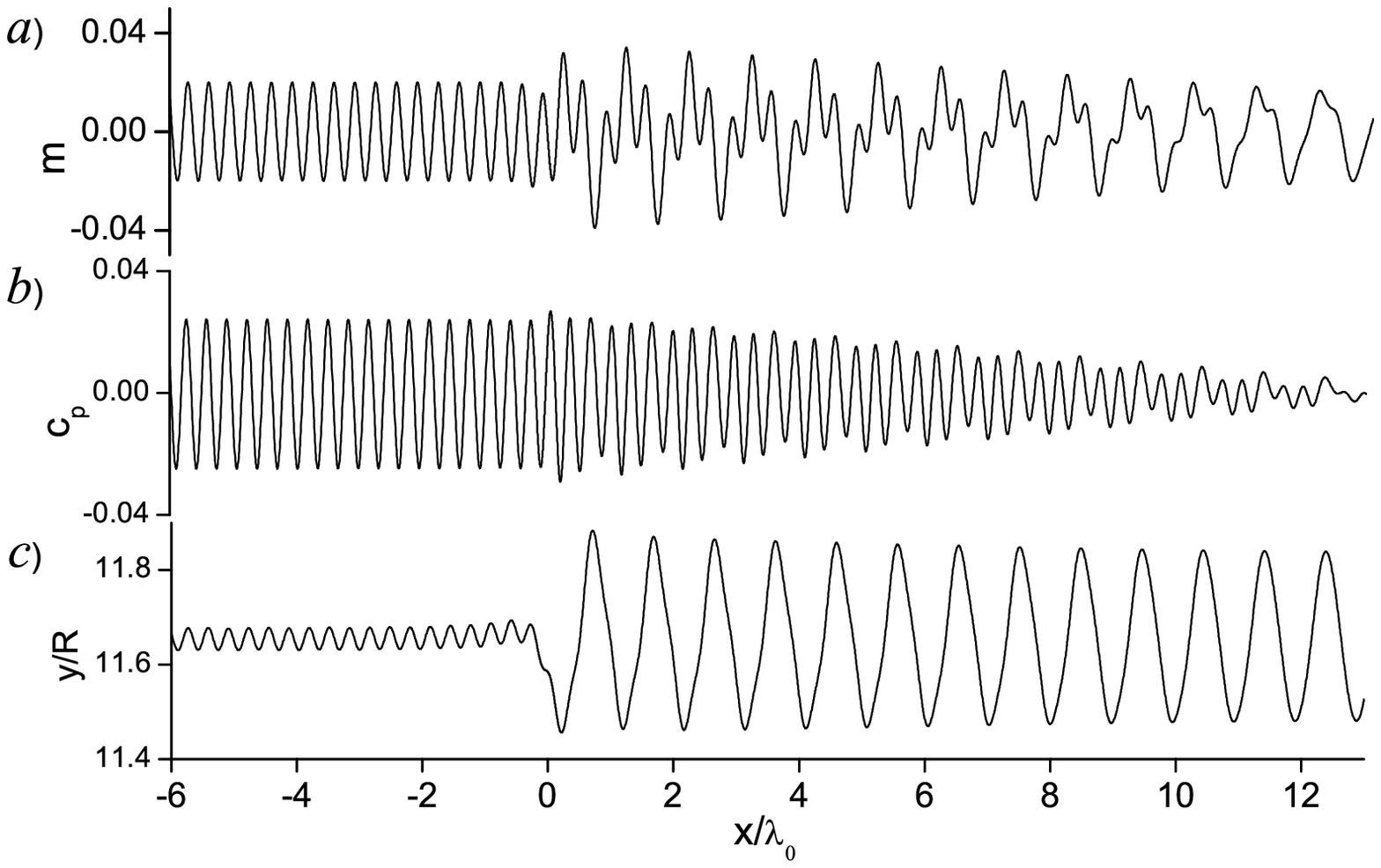}\\
\vspace*{-15mm}
\caption{Bending moment ($a$), pressure coefficient ($b$) and interface shape ($c$) at Froude number $F=2.0$ and the submergence $h=11.6$.}
\label{figure17}
\end{figure}
\begin{figure}
\centering
\includegraphics[scale=0.52]{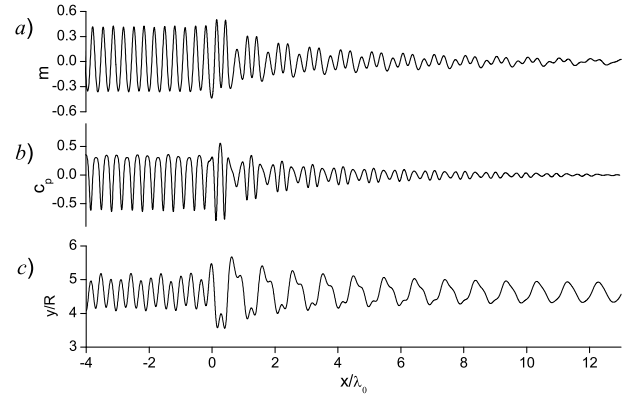}\\
\vspace*{-15mm}
\caption{The same as in Fig. 17 at the depth of submergence $h=4.61$.}
\label{figure18}
\end{figure}
The pressure coefficient and the bending moment along the interface are shown in Fig. \ref{figure17}$a,b$ for $h = 11.65$ and in Fig. \ref{figure18}$a,b$ for $h = 4.6$. They demonstrate interaction between the gravity and the elastic sheet waves. The wave due to the elastic sheet is dominating near the cylinder. It keeps constant amplitude in the upstream direction and gradually decays in the downstream direction. For $x/\lambda_0>8$, the amplitude of the bending moment in Fig. \ref{figure18}$a$ caused by the elastic sheet becomes smaller than that caused by the gravity wave, and the oscillations become qualitatively similar to that in Figs. \ref{figure13}$a$ and \ref{figure14}$a$. It is seen from Fig. \ref{figure18}$a$ that the distance at which the bending moment caused by the elastic sheet decays is much larger than that for $F=2.5$.
\begin{figure}
\centering
\includegraphics[scale=0.4]{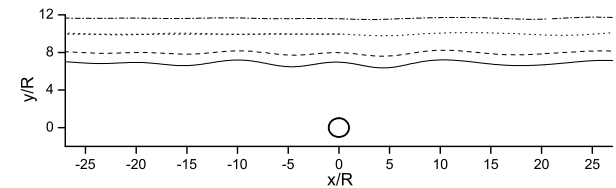}
\vspace*{-10mm}
\caption{Interface shape at different depths of submergence for Froude number $F=1.7$.}
\label{figure19}
\end{figure}
\begin{figure}
\centering
\includegraphics[scale=0.52]{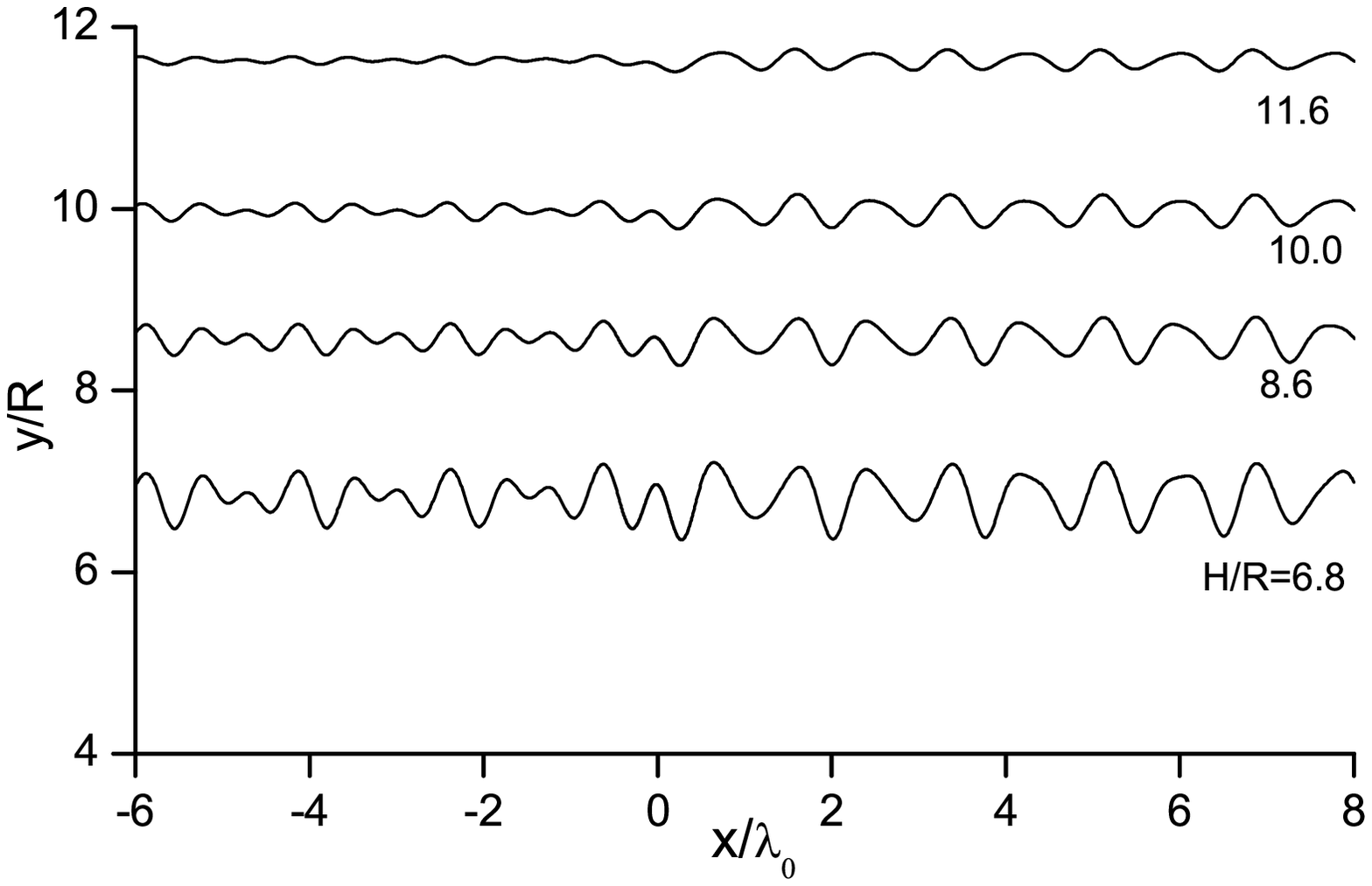}\\
\vspace*{-15mm}
\caption{Effect of submergence on the interface shape for the circular cylinder at Froude number $F=1.7$.}
\label{figure20}
\end{figure}
\begin{figure}
\centering
\includegraphics[scale=0.52]{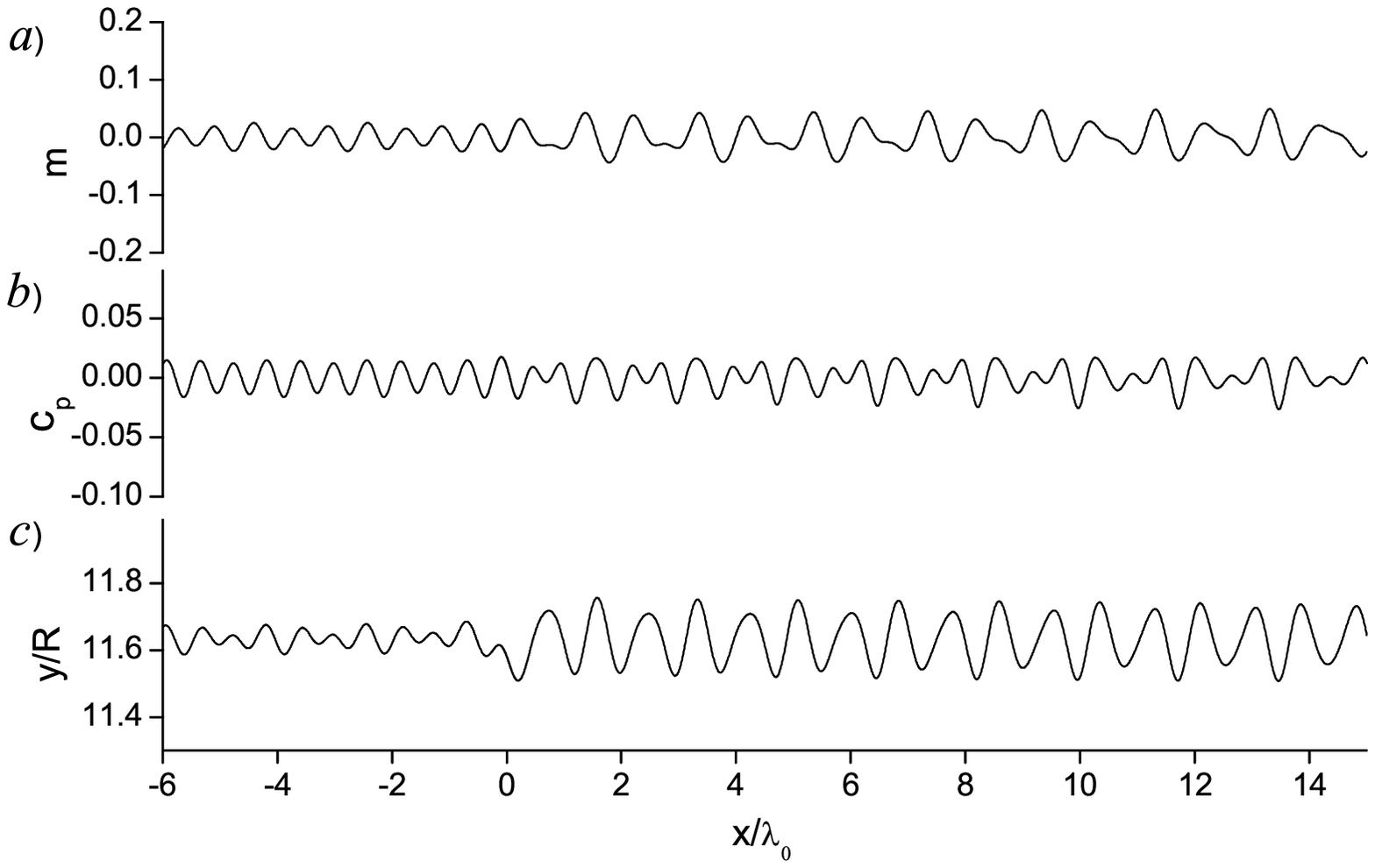}\\
\vspace*{-15mm}
\caption{Bending moment ($a$), pressure coefficient ($b$) and interface shape ($c$) at Froude number $F=1.7$ and the submergence $h=11.6$.}
\label{figure21}
\end{figure}
\begin{figure}
\centering
\includegraphics[scale=0.52]{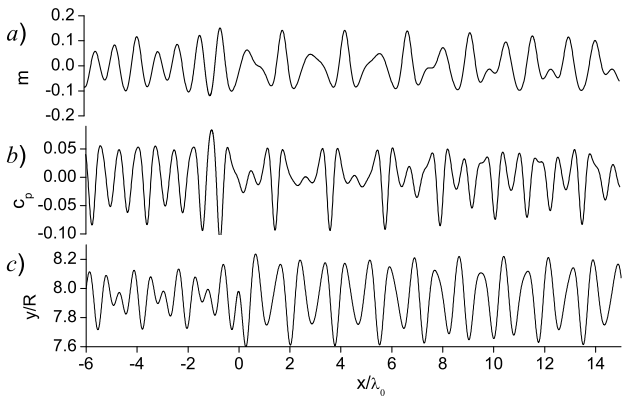}\\
\vspace*{-15mm}
\caption{The same as in Fig. \ref{figure21} at the depth of submergence $h=6.8$.}
\label{figure22}
\end{figure}
The results for Froude number $F=1.7$, which is quite close to the critical Froude number, $F/F_{cr}=1.03$, are shown in Figs. \ref{figure19}-\ref{figure22}. The wave numbers $k_{ice}$ and $k_w$ approach each other, and the ratio of the wave lengths becomes smaller, $\lambda_w/\lambda_{ice}$. The interfaces are shown in Figs. \ref{figure19} and \ref{figure20} for the depths of submergence from $11.6$ to the $6.8$. By comparing the interfaces at the depth $h=11.6$ for the Froude numbers $2.5$ and $2$ in Figs. \ref{figure12} and \ref{figure16}, respectively, we can find that the amplitude of the interface wave increases at the region upstream while the amplitude at the region downstream of the cylinder decreases. The latter agrees with the free-surface gravity flow past the submerged circular cylinder \cite{Sem_Wu2020}. The former indicates the larger effect of the elastic sheet at the smaller Froude number.

The shapes of the interface for different depths of submergence in Fig. \ref{figure20} are similar each other, but the amplitudes are different. In the upstream direction, $x/\lambda_0<0$, the shapes are periodic with period of about $2\lambda_0$, that corresponds to the superposition of two sinusoidal waves with the gravity and elastic sheet waves.

In the downstream direction, $x/\lambda_0>0$, the shape of the interface is not exactly periodic because the amplitude of the elastic sheet wave decays slowly downstream. By comparing the shape of the interfaces, behaviour of the bending moment and the pressure coefficient for different Froude numbers, we can see that the wave due to the elastic sheet decays  slower as the Froude number approaches the critical value. The similar behaviour of the bending moment and the pressure coefficient can be seen in Figs. \ref{figure21} and \ref{figure22}.

\begin{figure}
\centering
\includegraphics[scale=0.52]{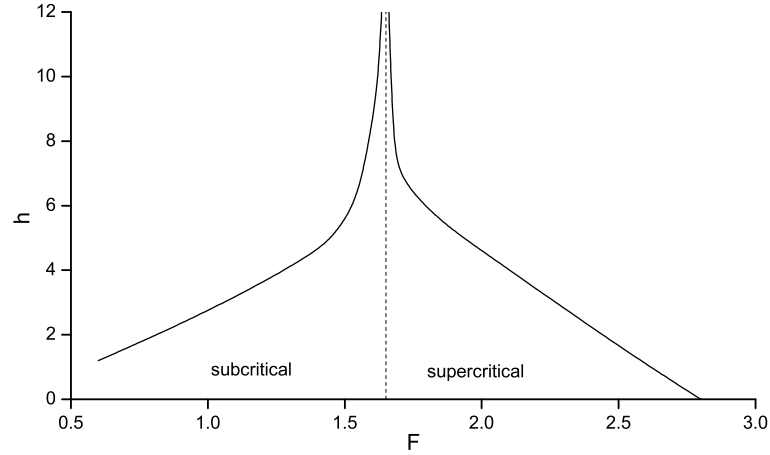}\\
\vspace*{-15mm}
\caption{Onset of the solution convergence.}
\label{figure22}
\end{figure}

The interaction between the fluid and elastic sheet and between the fluid and the submerged body is not always consistent for the steady flow. This situation has some analogy with the free-surface flow past the submerged cylinder, for which there is some combination of the Froude number and depth of submergence at which the steady solution does not exists. As the depth of submergence increases, the region of non-existence of the solution decreases and then disappears at large enough submergence. There is a maximal possible deflection of the free surface at which the dynamic boundary condition can be satisfied.
In the presence of the elastic sheet, the situation becomes more complicated because the dynamic boundary condition includes not only the deflection of the interface but also its derivatives. The term due to gravity in Eq. (31) increases as the Froude number decreases. It may cause such combination of the deflection, its derivatives and the velocity magnitude on the interface that the pressure distribution for the fluid (31) and for the elastic sheet (32) cannot be the same, or the dynamic boundary condition (4) cannot be satisfied.
For a supercritical flow, the fluid forces dominate the elastic forces, while for subcritical flows vice versa. This changes the flow configuration near the critical Froude number. The interface shown in Fig. 19 for F = 1.7 and   corresponds to the supercritical flow closest to the onset where the converged  solution can be obtained. It forms a hill over the cylinder, while for the subcritical flow in Fig. 6 for  , the interface forms a trough. The different limits of the flow configurations for sub- and supercritical regimes indicate an inconsistency of the fluid and elastic forces in some range near the critical Froude number. As the submergence increases, the deflection of the interface decreases as well as the range of Froude numbers at which the steady flow and the elastic sheet are not consistent. That can be seen in Fig. 23 where the onset of the region of existence of the steady solution is shown in the plane of the parameters Froude number vs. depth of submergence.

\section{\label{sec4} Conclusions}

A fully nonlinear numerical solution for the problem of steady gravity flow past a body submerged beneath an elastic sheet is presented in the form of the nonlinear analytical solution for the fluid part of the problem and the nonlinear Cosserat plate model applied to the elastic sheet, which are coupled throughout the numerical procedure. The solution of the fluid part of the problem is based on the integral hodograph method employed to construct the complex potential of the flow and Jacobi's elliptic theta functions to deal with the doubly connected fluid domain. The curvature and higher-order derivatives of the fluid boundary involved in the nonlinear Cosserat plate model have been evaluated using spline interpolation. The coupled problem has been reduced to a system of nonlinear equations with respect to the unknown magnitude of the velocity on the interface, which are solved using a collocation method.

Steady solutions of the full nonlinear problem were computed for sub- and supercritical regimes. For subcritical regimes, the dispersion equations have no real roots. The elastic sheet exhibits a most considerable deflection above the cylinder, which rapidly decays away. The deflection forms a curve symmetric about the Y-axis with a trough above the cylinder. The trough becomes larger as the depth of submergence decreases. These results qualitatively agree with linear solutions \cite{Savin_2012, Li_2019}. At the same time, the present nonlinear solution revealed a critical depth below which the deflection of the interface cannot provide balance between the bending and hydrodynamic forces in the steady flow. The critical submergence increases as the Froude number approaches the critical Froude number.

For supercritical regimes, the dispersion equations have two positive real roots which correspond to two wave numbers. The smaller wave number is closer to that corresponding to the gravity wave behind the cylinder without an elastic sheet, and the larger wave number appears in the presence of the elastic sheet. The dispersion equation does not restrict the flow regions in which the waves may occur, i.e. upstream or downstream of the cylinder. From the linear theories \cite{Savin_2012, Li_2019}, it was found that the gravity wave occurs downstream of the cylinder, while the wave due to the elastic sheet occurs upstream. The present nonlinear solution revealed that the waves may occur in both directions, but their amplitudes in each direction significantly depend on the perturbation of the interface and the ratio $F/F_{cr}$.

The calculations are presented for the thickness of the elastic sheet $b_i=0.05$ that corresponds to the critical Froude number $F_{cr}=1.65$. For larger thickness of the sheet, the critical Froude number increases. We expect that the flow configurations for larger thicknesses of the ice sheet will be similar to those for $b_i=0.05$ at the same ratio $F/F_{cr}$.

At high Froude number $F/F_{cr}>1.8$  and depth of submergence $h>6$, the interface shape is almost the same as it is without the elastic sheet. However, the effect of the elastic sheet can be seen in the behaviour of the bending moment and the pressure coefficient at the upstream. They oscillate with the wave number corresponding to the elastic sheet. The amplitude of oscillations is larger than that corresponding to the gravity wave at the downstream. At smaller submergence, the perturbation of the interface increases, and it becomes visible at the upstream. At very small depth of submergence, the elastic sheet starts to affect the whole interface: at the downstream, the amplitude of the gravity wave slightly decreases; and at the upstream, the gravity wave is excited in addition to the elastic sheet wave. The interface represents superposition of both these waves.

As the Froude number decreases and approaches the critical value $F_{cr}$, the wave caused by the elastic sheet can be observed in both directions. Its amplitude is constant in the upstream direction. In the downstream direction, the contribution of the elastic wave to the resulting shape decays. The closer the Froude number $F$  to the critical value $F_{cr}$, the slower decay is observed.

Similar to subcritical regimes, there is a critical submergence below which the steady supercritical solution cannot be obtained. The closer the Froude number to the critical value $F_{cr}$, the larger the critical depth of submergence. This may be caused due to the interaction of the two waves with closer wave length that may cause a resonance phenomenon.

\section*{DATA AVAILABILITY}

The data that support the findings of this study are available from the corresponding author upon reasonable request.

\end{document}